\documentclass[12pt, preprint]{aastex}
\def\deg{\ifmmode^\circ\else$^\circ$\fi}

\def\mic{~$\mu$m}

\def\mic{$\mu${\rm m}}
\def\mujy{$\mu${\rm Jy}}

\def\arcs{\ifmmode {''}\else $''$\fi}
\def\arcm{\ifmmode {'}\else $'$\fi}
\def\parcs{\sa=.07em \sb=.03em
    \ifmmode $\rlap{.}$^{\scriptscriptstyle\prime\kern -\sb\prime}$\kern -\sa$
    \else \rlap{.}$^{\scriptscriptstyle\prime\kern -\sb\prime}$\kern -\sa\fi}
\def\parcm{\sa=.08em \sb=.03em
    \ifmmode $\rlap{.}\kern\sa$^{\scriptscriptstyle\prime}$\kern-\sb$
    \else \rlap{.}\kern\sa$^{\scriptscriptstyle\prime}$\kern-\sb\fi}

\def\spose#1{\hbox to 0pt{#1\hss}}
\def\simlt{\mathrel{\spose{\lower 3pt\hbox{$\mathchar"218$}}
    \raise 2.0pt\hbox{$\mathchar"13C$}}}
\def\simgt{\mathrel{\spose{\lower 3pt\hbox{$\mathchar"218$}}
    \raise 2.0pt\hbox{$\mathchar"13E$}}}
\def\lsim{\rlap{$<$}{\lower 1.0ex\hbox{$\sim$}}}
\def\gsim{\rlap{$>$}{\lower 1.0ex\hbox{$\sim$}}}

\begin{document}

\title{Spitzer$^1$\ IRS$^2$\ 16 micron Observations of the GOODS
 Fields }

\altaffiltext{1}{Based on observations obtained with the {\it Spitzer
    Space Telescope}, which is operated by JPL, California Institute
  of Technology for the National Aeronautics and Space
  Administration}

\altaffiltext{2}{The IRS is a collaborative venture between Cornell
University and Ball Aerospace Corporation that was funded by NASA
  through JPL.}

\author{Harry I. Teplitz\altaffilmark{3}, 
Ranga Chary\altaffilmark{4},
David Elbaz\altaffilmark{5}, 
Mark Dickinson\altaffilmark{6},
Carrie Bridge\altaffilmark{7},                                  
James Colbert\altaffilmark{8},
Emeric Le Floc'h\altaffilmark{5},
David T. Frayer\altaffilmark{9},
Justin H. Howell\altaffilmark{7}, 
David C. Koo\altaffilmark{10},
Casey Papovich\altaffilmark{11},
Andrew Phillips\altaffilmark{10},
Claudia Scarlata\altaffilmark{8}, 
Brian Siana\altaffilmark{7}, 
Hyron Spinrad\altaffilmark{12}, and
Daniel Stern\altaffilmark{13}                                
}

\altaffiltext{3}{Infrared Processing and Analysis Center, MS 100-22, Caltech, Pasadena, CA 91125.  hit@ipac.caltech.edu}
\altaffiltext{4}{MS220-6, US Planck Data Center, Caltech, Pasadena, CA 91125}
\altaffiltext{5}{CEA-Saclay, DSM/DAPNIA/Service d'Astrophysique, 91191 Gif-sur-Yvette Cedex, France}
\altaffiltext{6}{National Optical Astronomy Observatory, 950 North Cherry Street, Tucson, AZ 85719, USA}
\altaffiltext{7}{Division of Physics, Math, and Astronomy, California Institute of Technology, Pasadena, CA 91125}
\altaffiltext{8}{Spitzer Science Center, MS 220-6, Caltech, Pasadena, CA 91125}
\altaffiltext{9}{NRAO, PO Box 2, Green Bank, WV  24944}
\altaffiltext{10}{Department of Astronomy and Astrophysics, University of California
Observatories/ Lick Observatory, University of California, Santa Cruz, CA 95064.}
\altaffiltext{11}{Department of Physics and Astronomy, Texas A\&M
  University, College Station, TX 77843-4242}
\altaffiltext{12}{Department of Astronomy, University of California at Berkeley,
Mail Code 3411, Berkeley, CA 94720.}
\altaffiltext{13}{Jet Propulsion Laboratory, California Institute of Technology, Pasadena, CA 91109}

\begin{abstract}

 We present Spitzer 16 micron imaging of the Great Observatories
 Origins Deep Survey (GOODS) fields. We survey 150 square arcminutes
 in each of the two GOODS fields (North and South), to an average 3 sigma depth
 of 40 and 65 \mujy\ respectively.  We detect $\sim 1300$\ sources in both fields combined. We
 validate the photometry using the 3-24 \mic\ spectral energy distribution of stars in the fields
 compared to Spitzer spectroscopic templates.  Comparison with ISOCAM
 and AKARI observations in the same fields show reasonable
 agreement, though the uncertainties are large.  We provide a catalog
 of photometry, with sources cross correlated with available Spitzer,
 Chandra, and HST data.  
Galaxy number counts
 show good agreement with previous results from ISOCAM and AKARI,
 with improved uncertainties.  We examine the 16 to 24 \mic\ flux
 ratio and find that for most sources it lies within the expected
 locus for starbursts and infrared luminous galaxies.  A color cut of
 $S_{16}/S_{24}>1.4$\ selects mostly sources which lie at
 $1.1<z<1.6$, where the 24 \mic\ passband contains both the 
redshifted 9.7 \mic\ silicate absorption and the
 minimum between PAH emission peaks.  We measure the integrated galaxy light of 16 \mic\
 sources, and find a lower limit on the galaxy contribution to the extragalactic
background light at this
 wavelength to be $2.2\pm 0.2$\ nW m$^{-2}$\ sr$^{-1}$.

\end{abstract}


\section{Introduction}

UV light absorbed by dust is primarily reradiated in the far-IR (FIR),
with a peak between 60 and 100 \mic. In addition, complex molecules,
the polycyclic aromatic hydrocarbons (PAHs), radiate characteristic
emission features in the mid-IR (MIR), the most prominent of which are
at wavelengths 6.2, 7.7, 8.6, 11.3 and 12.7 \mic\ \citep[see][for a
review]{1989ARA&A..27..161P}. Over the same wavelength range, there is
continuum emission from very small dust grains, which can dominate PAH
emission at wavelengths beyond 10 \mic\ \citep{2000A&A...359..887L}.
The MIR flux, which results from the sum of these two emission
mechanisms, correlates strongly with the integrated IR luminosity from
8-1000 \mic, $L_{\rm IR}$, which is a direct tracer of star formation
\citep[][and references
therein]{1998ARA&A..36..189K,2001ApJ...556..562C}. Active galactic
nuclei (AGN) are also strong MIR sources, making the infrared an
excellent tracer of obscured AGN which may not be accessible even to
ultradeep X-ray observations
\citep{2004ApJS..154..166L,2005ApJ...621..256S,2005ApJ...631..163S,
  2007ApJ...670..173D,2008ApJ...687..111D}. Observations from the {\it
  Infrared Space Observatory}\ \citep[ISO; ][]{1996A&A...315L..27K},
the {\it Spitzer
 Space Telescope}\ \citep{2004ApJS..154....1W}, and the 
AKARI satellite \citep{2007PASJ...59S.369M} have revolutionized the
study of infrared luminous sources in the past decade.

Of particular interest is the measurement of the integrated galaxy light
and its value compared to the DIRBE measured extragalactic background light
\citep[EBL; for a review, see][]{2005ARA&A..43..727L}. 
Lower limits on the EBL are inferred from the detection of galaxies in imaging
surveys with, e.g. ISO and Spitzer \citep{2002A&A...384..848E,
 2006A&A...451..417D}. Upper limits on the background are obtained
from $\gamma$-ray observations of blazars, because of the absorption
of TeV photons by the cosmic infrared background through pair production
\citep[e.g.][]{2002A&A...384L..23A}. More recently, \cite{2010arXiv1002.3674M}\ have used
long-wavelength (65-160 \mic) AKARI imaging to directly detect the
cosmic infrared background, and confirm a value in excess of the lower limit
measured by stacking of Spitzer sources.  Considerable effort has gone into
predicting the contribution of AGN to the EBL \citep[][among others]
{2002A&A...383..838F,2006ApJ...642..126B,2006ApJ...640..603T,2007ApJ...660..988B}.
It is clear that galaxies hosting AGN do not dominate (well under
30\%), but their precise contribution remains hard to estimate, in
part due to the trouble identifying obscured sources. Measurement of
the resolved portion of the background at a particular wavelength is
performed by integrating the total flux from individually
detected galaxies and through the technique of stacking.

Deep, MIR observations of the Great Observatories Origins Deep Survey
\citep[GOODS;][]{2004ApJ...600L..93G,2003mglh.conf..324D} fields have
led to important advances in understanding the global history of star
formation and AGN evolution.  For example, the peaks in the
differential source counts at 15 and 24 microns at 0.4 and 0.2 mJy,
respectively
\citep{2002A&A...384..848E,2004ApJS..154...80C,2004ApJS..154...70P,2004ApJS..154...66M}, 
reflect the contribution of luminous
infrared galaxies (LIRGs) at higher redshifts
\citep[e.g.][]{2005MNRAS.358.1417P}\ as MIR features redshift into the
bands.  The GOODS fields also provide substantial information 
on the presence of AGN, both obscured and unobscured, using the
ultradeep Chandra observations
\citep{2001AJ....122.2810B,2002ApJS..139..369G}.  

In this paper, we present Spitzer 16 \mic\ observations covering the
GOODS fields. We provide the source catalog, including quality
assessment flags, and provide source associations between these
observations and other available photometry from Spitzer, HST, and
Chandra. We discuss preliminary analysis of the dataset including
number counts, 16/24 \mic\ colors, and the integrated galaxy light of 16 \mic\
selected sources (including obscured AGN). Earlier Spitzer
imaging of a small region within GOODS-North was presented in
\cite{2005ApJ...634..128T}.

\cite{2009arXiv0901.3783L}\ used an earlier reduction of the present
Spitzer survey as part of a study of the cosmic star-formation history
using MIR number counts.  They combine a non-parametric inversion of
galaxy counts at 15--850 \mic\ with constraints from measurements of
the cosmic infrared background. 
They exclude a major contribution from
``hyper-LIRGs'' at high redshift, concluding that these sources may in
fact be AGN-dominated.  In addition, \cite{2009PASJ...61..177B}\
recently reported imaging of a portion of GOODS South at 15 \mic\
using the Infrared Camera (IRC) onboard AKARI.  Their analysis
has some overlap with the Spitzer study (number 
counts, MIR colors), and the results are consistent, as we discuss below.
While the two surveys are complementary, the Spitzer study had the advantage
of significantly more telescope time; it covers more area and to greater 
depth, especially with the addition GOODS-North.  We report $\sim 1300$\ objects
over the two fields, compared to $<300$\ in the AKARI survey.

We describe the survey, source extraction, quality assessment, and
validation in Section 2.  We provide the catalog and discuss the
survey properties in Section 3.  In Section 4 we describe preliminary
analysis of the data, before summarizing in Section 5.  Throughout, we
assume a $\Lambda$-dominated flat universe, with $H_0=71$\ km
s$^{-1}$\ Mpc$^{-1}$, $\Omega_{\Lambda}=0.73,$\ and $\Omega_{m}=0.27$.

\section{Observations and Data Reduction}

In this section we describe the observations and data reduction.  In
addition, we have performed several validation checks and quality
assessment measurements, which we describe in detail.  

The {\it Spitzer}\ peak-up imaging (PUI) capability offered 
a sensitive imaging capability at 16 \mic\ using the Infrared
Spectrometer
\citep[IRS;][]{2004ApJS..154...18H}. The peak-up array was read
out on the same detector as the Short-Low (5-14.5 \mic) spectroscopic channel, in a
small field of view (56\arcsec $\times$\ 80\arcsec).  The Si:As 
detector was similar to that used in the 24 \mic\ channel of the MIPS
instrument \citep{2004ApJS..154...25R}. In 10 minutes of observation, the PUI 
achieved 5$\sigma$\ depths of $\sim 45~\mu$Jy.

Data were taken in two general observer {\it Spitzer}\ programs:
GO-3661, observing GOODS-South in Cycle 1; and GO-20599, observing
GOODS-North in Cycle 2.  The southern survey was taken before the PUI
mode was fully commissioned for {\it Spitzer}.  In order to observe
GOODS-South at 16 \mic, the instrument was commanded to take a series
of short spectroscopic observations over a range of positions, and the
images were acquired in parallel.  Similar observations were taken over
a small area within GOODS-North and are described in
\cite{2005ApJ...634..128T}.  The northern survey used the standard PUI
mode.

The southern survey was designed to consist of $\sim 130$\ square
arcminutes of shallow data (2 minutes per pointing) and 10 square
arcminutes of deeper data (16 minutes per pointing) in the area of the
{\it Hubble}\ Ultradeep Field \citep[UDF; ][]{2006AJ....132.1729B}.
Observations were first taken in early 2005, but were compromised by
persistent charge on the detector resulting from the preceding
program, which targeted the rings of Jupiter.  Much of the survey was repeated later in the year.  Many
exposures within the compromised observations were unaffected, so the
shallow survey has an average of 4 minutes per pointing, with
small areas of overlapping frames having greater depth. The UDF coverage varies
from 16 to 32 minutes.

The northern survey covered 150 square arcminutes with $\sim 10$\ minutes per
pointing, observed in 2006.  We chose not to combine these data
with the previous GOODS North imaging \citep{2005ApJ...634..128T}, given
the small area of the latter and the different observing modes.

IRS 16 \mic\ images were reduced by version S13.2 of the standard Spitzer Science
Center (SSC)
pipeline\footnote{http://ssc.spitzer.caltech.edu/irs/dh/}.  The
pipeline supplied Basic
Calibrated Data (BCD) frames with most instrumental
effects corrected and flux calibration applied.  Flux calibration was
updated to match the latest version of
the pipeline (S18.7).  
The pipeline removed a nominal low-background
sky image, but some residual zodiacal light could have remained.  We created
median sky images from near-in-time subsets of the data, specifically from
each contiguous block of observations or ``astronomical observing request'' (AOR).
We then subtracted the median sky frames after scaling to the mode of the images.  

Individual PUI frames are quite small (56\arcsec\ by 80\arcsec),
as the primary purpose of the 16 \mic\ camera was target acquisition.
The plate scale is $\sim 1.8$\ arcseconds per pixel.  Geometric
distortion is about 2\%.

\begin{figure}[t!]
\plottwo{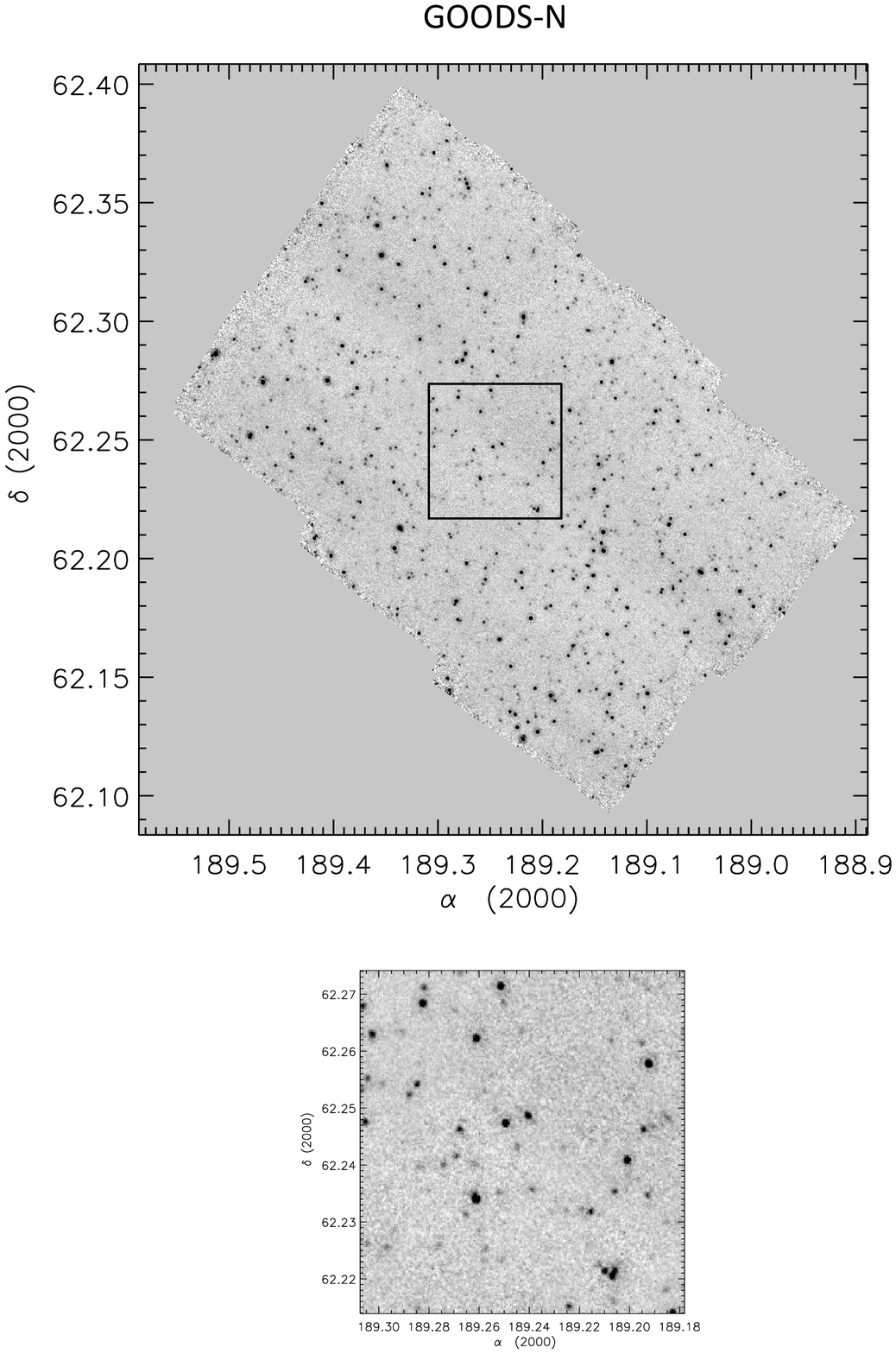}{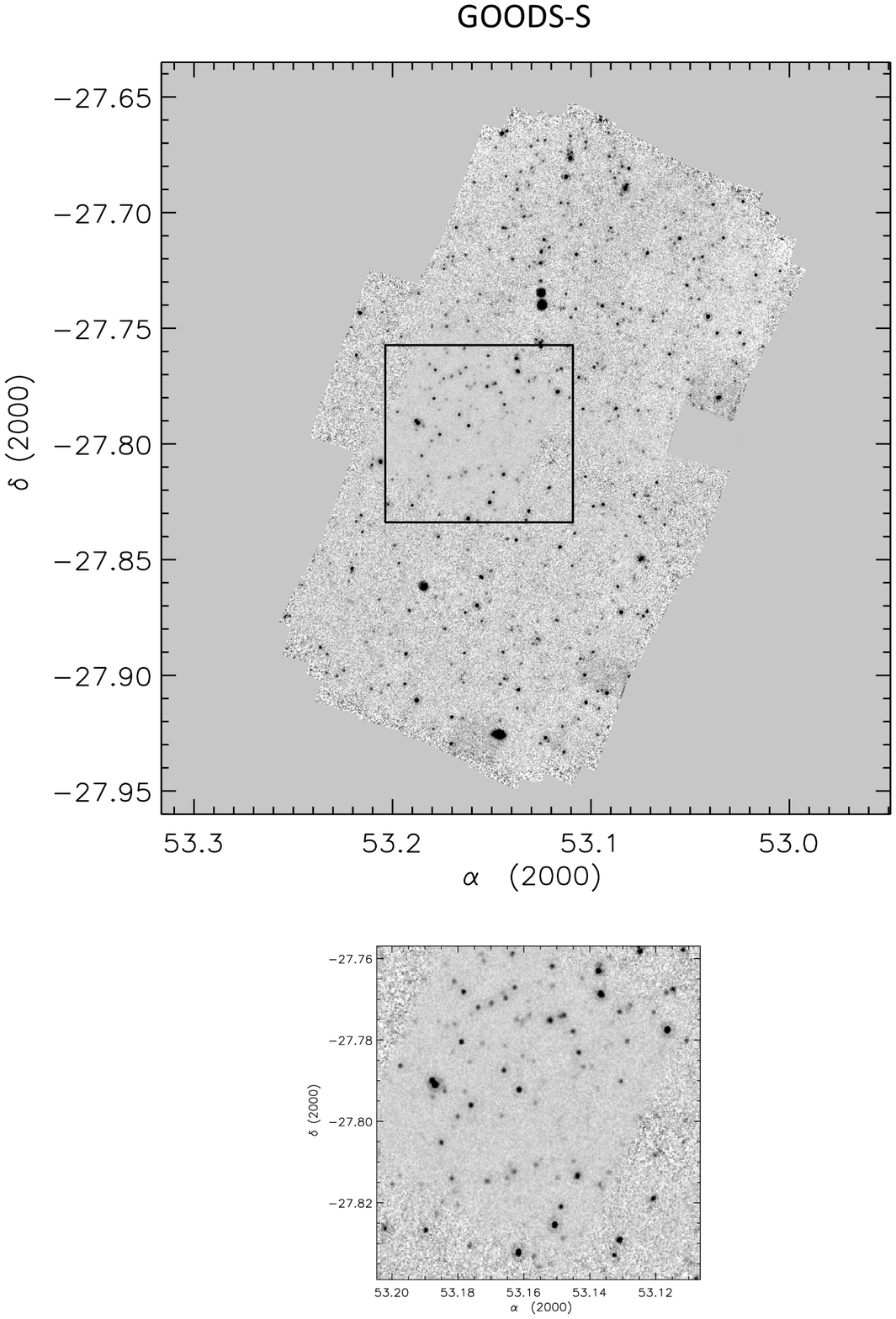}
\caption{\label{fig: image} We show the mosaic of the North (left) and South
(right) GOODS fields.  A smaller region is shown below to give a better indication
of crowding and image quality;  for the South, the region includes the
UDF, which has greater sensitivity than the rest of the field.}
\end{figure}

We registered and combined images using the MOPEX software distributed by
the SSC
\citep{2005PASP..117.1113M, 2006SPIE.6065..330M, 2006SPIE.6274E..10M}.
We employed {\it drizzle}\ interpolation \citep{2002PASP..114..144F}.
MOPEX drizzle produces the same results as {\it wdrizzle} in
IRAF\footnote{IRAF is distributed by NOAO, which is operated by AURA
 Inc., under contract to the NSF}.
The package uses the World
Coordinate System (WCS) definition of both spatial offsets and
geometric distortion.  In imaging mode, the IRS did not perform an
initial peak-up to refine the pointing, so the absolute WCS is good to
only about 1 arcsecond, though the pointing is considerably better between
BCDs within a single AOR.  The Point Spread Function (PSF) at
16 \mic\ has a full width at half maximum (FWHM) of about 3.6
arcseconds.  The final mosaics have a plate scale of 0.9\arcs/pixel,
and the {\it pixfrac}\ was set to 0.6.  Figure \ref{fig: image}\ shows the
mosaics of the GOODS fields.

Photometry was performed by applying a custom point source extraction
code which utilizes PSF-fitting and positional priors
\citep{2004ApJS..154...80C}. We used 5$\sigma$\ sources from the full
GOODS IRAC channel 1 catalog (3.6 \mic; Dickinson et al.\ in
preparation) as input. Sources within a 20\arcs$\times$ 20\arcs\ box
around each input source were simultaneously fit to the measured PSF.
The same technique was used in the preparation of the GOODS 24 \mic\
catalog (Chary et al., in preparation). The PSF for each field (North
and South) was measured by registering about 15 bright point sources
in each field. Source position was allowed to vary by up to an
arcsecond to account for the uncertainty in the WCS. In addition to
pointing issues, this flexibility has other advantages as well, 
including the possibility of actual displacements between the
centroids of the starlight (IRAC) and the dust emission (MIPS or IRS),
or cases where IRAC sources are blended (leading to a displacement of
the centroid), but where one source may dominate the MIPS or IRS
emission. We checked the residual map to ensure that no sources were
missed by using the IRAC priors.

The flux uncertainty was measured from 
residual pixels after subtracting the best-fit estimate of the source;
the variance was taken as the PSF weighted sum of squares of the residuals.
However, the noise in the drizzled images is correlated due to the
sub-sampling. To correct for this effect, we scaled the measured
uncertainty upwards by a factor of $\sim 1.7$\ \citep[see
][]{2000AJ....120.2747C}.

Photometric measurements are reported in Tables 1 and 3.  We
report the position of each object as that of the IRAC positional
prior which was used as input to the source extraction.  We cross
correlated these positions with those of objects in the IRAC and MIPS
24 \mic\ catalogs (Dickinson in prep. and Chary in prep.,
respectively) and report their photometry as well.  The IRAC photometry
was measured with SExtractor \citep{1996A&AS..117..393B}, using aperture photometry except in the
case of a small number of extended sources for which MAG\_AUTO is used.

In general, we discarded sources with less than 5$\sigma$\ significance. This
uniform cut allows us to consider the entire survey area, despite its
non-uniform coverage. Final fluxes were consistent with aperture
photometry, with appropriate aperture corrections. As described below,
a small number of additional sources are reported in the catalog with
less than 5$\sigma$\ significance (with a corresponding quality flag),
because they were expected to meet that criterion based upon the
integration time.

Furthermore, to ensure a low incidence of spurious sources, we only used
areas of the survey with good coverage.  In the North and the UDF, we required
at least ten individual data collection events (DCEs) per pixel.  In the
shallow Southern survey, this depth was not possible, but we did reject
areas with less than 2 DCEs coverage.  Depth of coverage is reported
in the table.

The IRAC observations of the GOODS fields are much deeper than the
current survey. The 3.6 \mic\ catalog was used as input positional
priors, but we can use the 8 \mic\ data as a quality check. Even
accounting for the steep slope of IR luminous galaxies, it is unlikely
that real sources will be detected at 16 \mic\ and not at 8 \mic.
We would reject as spurious any source which is not reported at
$5\sigma$\ significance in the IRAC channel four catalog. However, all
sources which meet our other selection criteria pass this test.
Further quality assessment is discussed below.

\subsection{Quality Assessment}

We performed several quality assessment procedures in order to flag objects
which may be less reliable.  Flags are reported in the table.

First, we created a residual map by scaling the PSF to the measured fluxes
and subtracting it from the image mosaic.  We identified a small number of
bright, extended objects (9 in the North and 7 in the South) for which the
PSF-fit photometry is inappropriate.  We measured these sources in a 
large aperture instead.  A flag for extended sources is reported in the table.

Second, we estimated the concentration of detected objects by measuring
the ratio of aperture photometry in 6 and 2 pixel radii.  Objects which
are point sources should fall in the range of $\sim 2-3:1$. 
Objects outside the nominal range (ratio $>4$\ or $<1.3$) are flagged in the catalog.
Several objects with anomalous values were inspected by eye and rejected
from the catalog.

Third, we flagged objects which may be affected by source confusion.  We
identified close companions in the catalog, within a radius of 5.4
arcseconds (6 pixels, a little larger than the FWHM of the PSF). 
We
also flagged targets which have 8 \mic\ sources (detected at $>3\sigma$)
within 4 arcseconds,
whether or not they are detected at 16 \mic.

Fourth, we compared the signal to
noise ratio (SNR) estimate from our photometry code to the
expectation from the exposure time map.  The exposure time calculator
at the Spitzer Science Center website predicts a 1$\sigma$\
sensitivity of $\sim 50$\ \mujy\ in a single 30 second exposure, and
the observations are background limited so sensitivity is expected to
scale with the square root of exposure time.  The measured SNR 
generally agrees within about 20\% with
expectations, though our estimates are slightly conservative.  A small
number of objects were selected as meeting the 5 sigma cut but exceeding
the expected sensitivity; that is, they would have been excluded using
the expected noise values. We retain these objects and flag them as
long as the expected SNR is 4 or greater. Conversely, a somewhat
larger number of objects were expected to meet the 5 sigma cut but did
not in our measurements.  We report such objects as low confidence
sources as long as they have measured SNR greater than 4.

Finally, we flagged sources with the spectral energy distribution (SED) expected for stars.
Specifically, we identified stars as those objects for which the flux
density is declining across all six bands. We exclude one object for which $f_{24}>f_{8}/2$, which
can be the case for some low redshift galaxies.  Of these, 7 (2) in the N (S) are confirmed as
stars in ground-based spectra. In the North, we also flag as stars
two sources that are saturated in IRAC channel 1 (which causes them to fail the
SED check) but are confirmed in ground-based spectra.  We check all star identifications
in the ACS imaging of the GOODS fields and confirm that they are point sources.  The
imaging also shows that two of the objects
identified as stars in the North are likely confused with nearby galaxies, and we 
flag these separately (see Section \ref{sec: results}).

\subsection{Photometric Calibration and Verification}

The calibration of the Spitzer Peak-Up Imaging mode is in 
the IRS Instrument Handbook\footnote{http://ssc.spitzer.caltech.edu/irs/irsinstrumenthandbook/}
and briefly summarized here.  The current calibration is based upon
spectroscopic and imaging observation of four A stars for which
detailed spectral models were available.  There is a systematic uncertainty of 
$\sim 5$\% in the calibration.

Stellar fluxes were measured from the the IRS spectra of the
calibrators by integrating under the filter transmission curve and
color-correcting to the effective wavelength (15.8 $\mu$m) assuming
$\nu F_{\nu}=const$.  This assumption is designed to keep the color
correction small for a wide range of spectral slopes.  The calibration
is normalized to infinite aperture using Tiny Tim
V2.0\footnote{http://ssc.spitzer.caltech.edu/archanaly/contributed/stinytim/index.html}.
In practice, aperture photometry requires an aperture correction.
Profile-weighted fitting, such as that utilized in the present study,
is normalized to a finite radius and then aperture corrected as well.

\begin{figure}[t!]
\plotone{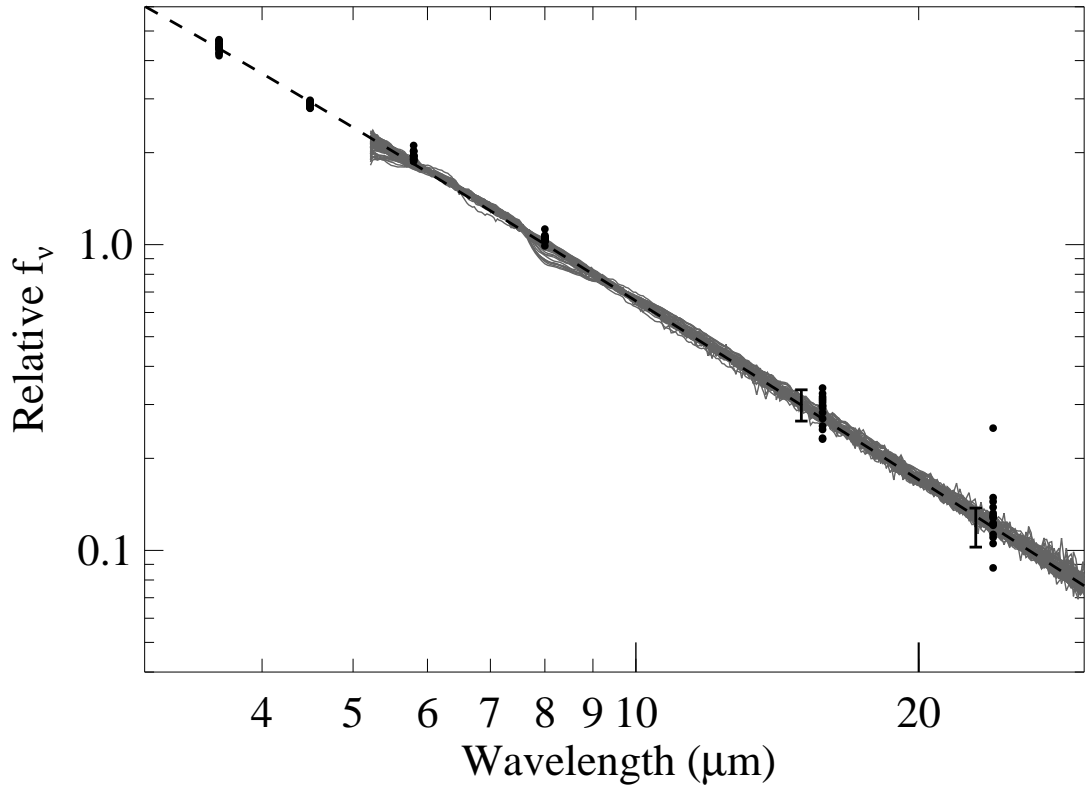}
\caption{\label{fig: checkstars} Spitzer photometry
for 15 stars in GOODS (circles), normalized to
the the average of the 6-band photometry.  Data are compared to IRS spectra of stars
with stellar types A through M (grey spectra) and
the \cite{1979ApJS...40....1K} model for an A0V star (dashed 
spectrum). The IRS spectra are taken from the SASS survey (Ardila et al. in prep). 
Error bars are small for the IRAC data
points; typical error bars for the 16 and 24 \mic\ photometry are shown
next to the data.  }
\end{figure}

The PUI calibration is tied to the calibration of the spectrometer,
which has been validated against a large number of stars.  To evaluate
the calibration in the particular case of the GOODS data, we can
compare the photometry of stars within the two fields.  Twenty-five
objects have IRAC colors clearly indicative of stars.  However, four
of these objects have quality flags greater than 1, so we exclude
them.  We also exclude another 6 objects that have SEDs indicating 
possible IR excesses.  So, we have 15 stars with which to validate the
PUI calibration.  These objects are faint, however, and we do not know
their stellar types.  Nonetheless, the mid-infrared spectral energy
distribution of stars does not vary much across a wide range of
stellar types.  The Spitzer Atlas of Stellar Spectra (SASS, Ardila et
al. in preparation) observed stars across many stellar types with the
IRS.  In Figure \ref{fig: checkstars}\, we plot the photometry for the
15 stars compared to SASS IRS spectra of stars with types A through M,
excluding super giants, and the \cite{1979ApJS...40....1K} model for
an A0V star.  We apply a small ($<4$\%) color-correction to the
photometry to account for the difference between the assumed
calibration reference spectrum ($\nu f_{\nu}=const$\ for 
IRAC\footnote{IRAC Instrument Handbook; http://ssc.spitzer.caltech.edu/irac/iracinstrumenthandbook/} 
and PUI;
$10^4$\ K black body for 
MIPS\footnote{MIPS Instrument Handbook; http://ssc.spitzer.caltech.edu/mips/mipsinstrumenthandbook/}) and a typical (5000 K black body)
stellar spectrum.  The spread in photometry is similar to the expected
variation within stellar types, and does not indicate systematic
offsets in the photometric calibration.

\subsection{Comparison with ISOCAM and AKARI}

In addition to validation using stars, we compare {\it Spitzer}
photometry to other observations of the field at similar wavelengths.

\begin{figure}[t!]
\plotone{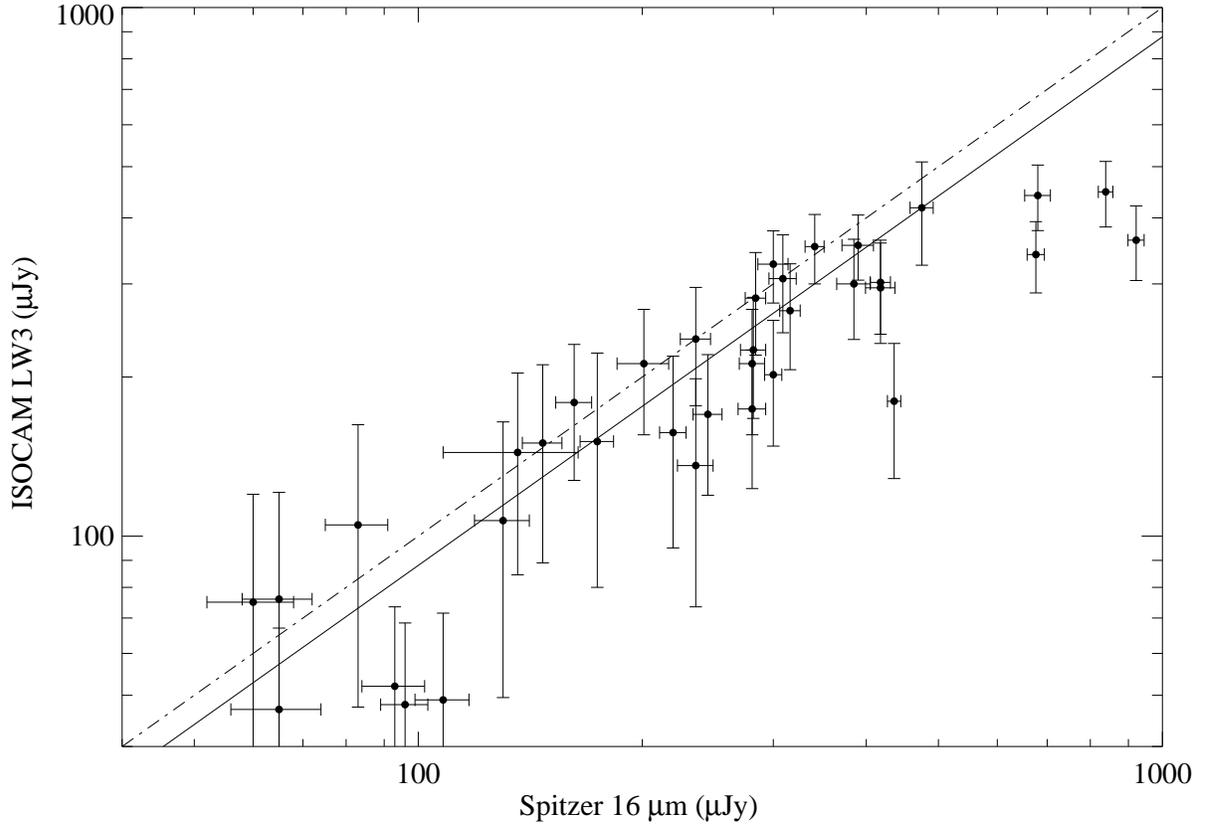}
\caption{\label{fig: isocam} 
The flux density measured with 
{\it Spitzer} PUI in the 16 \mic\ bandpass compared to that measured
with ISOCAM in the 15 \mic\ bandpass \citep[LW3][their Table 3]{1999A&A...342..313A} for objects 
detected by both instruments.  The {\it dot-dashed} line indicates equal flux, and the {\it solid} line
indicates the best fit offset of a factor of 1.13. 
}
\end{figure}

ISOCAM \citep{1996A&A...315L..32C} on-board the {\it Infrared Space
 Observatory}\ observed a portion of
GOODS-North centered on the Hubble Deep Field North
\citep{1996AJ....112.1335W}. Careful reduction of those data was
presented by \cite{1999A&A...342..313A}. The ISOCAM catalog reports 40
objects detected at 15 \mic\ in the main catalog. Of these, 36 are
associated with sources in the Spitzer 16 \mic\ catalog. The four
unmatched sources are all fainter than 100 \mujy, and two of them do not appear in
the (deeper) 24 \mic\ catalog, indicating that they are probably spurious. 
Figure \ref{fig: isocam}\ shows the comparison of Spitzer and ISOCAM fluxes for
matched sources. 

Both ISOCAM and Spitzer PUI are calibrated assuming
$\nu f_{\nu} = const$, but the effective wavelengths are
substantially different, being 14.3
and 15.8 \mic, respectively\footnote{See the ``ISOCAM
 Photometry Report'', 1998;
 http://www.iso.vilspa.esa.es/users/expl\_lib/CAM/photom\_rep\_fn.ps.gz}.
So, a color-correction to the ISOCAM photometry equal to the
ratio of the effective wavelengths -- e.g., a factor of
1.1 -- is expected.  We fit the offset between Spitzer and ISOCAM photometry assuming a
constant slope of unity and find Spitzer fluxes to be 1.13 times brighter.
Overall, we see general agreement within the large error bars. Spitzer
had better spatial resolution at 16 \mic\ than ISOCAM, and some of the
ISOCAM sources are blends of multiple objects, and this may explain
the difference in flux for some objects. The four
brightest objects do not appear to suffer from blending, however. 

\begin{figure}[t!]
\plotone{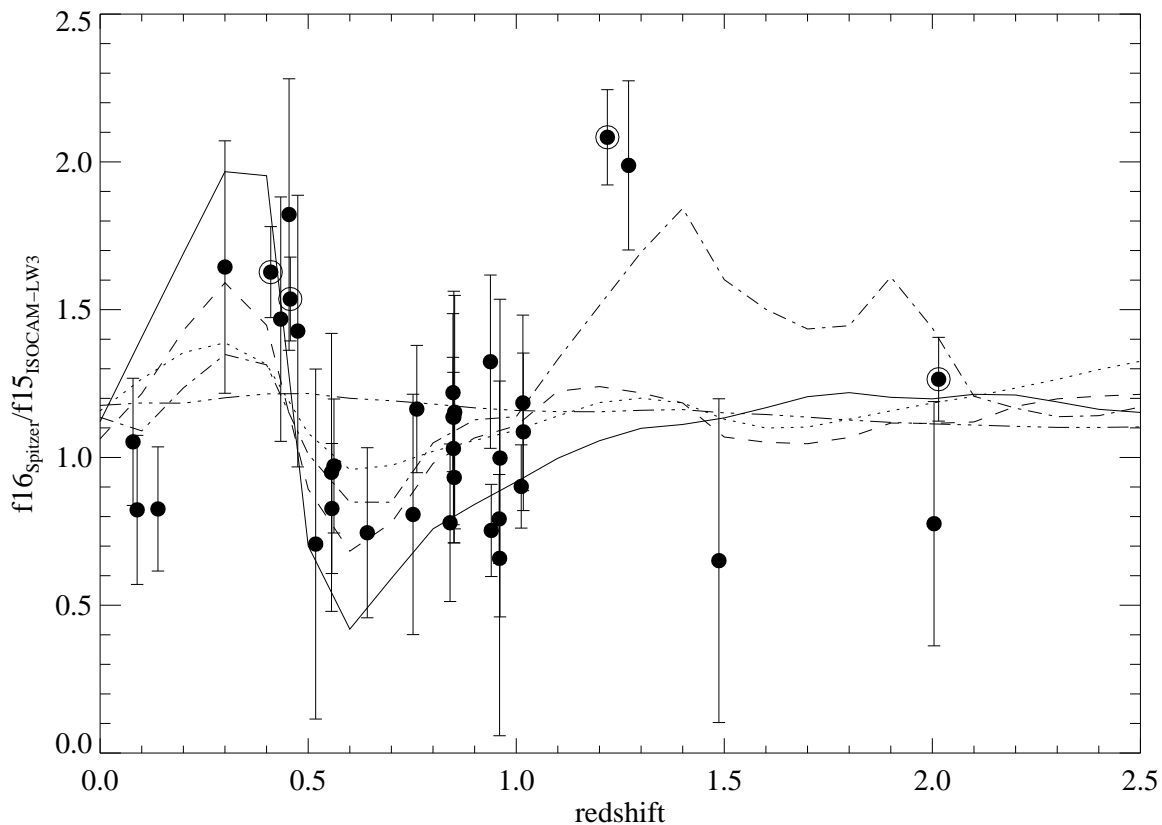}
\caption{\label{fig: iso_ratio} 
The ratio of the Spitzer 16 \mic\ to ISOCAM 15 \mic\ (LW3) flux
densities as a function of spectroscopic redshift. The four brightest
sources are circled.  Also plotted
are predicted ratios based on Spitzer IRS spectra of template galaxies:
the extreme silicate-absorption galaxy
 IRAS F00183-7111 \citep[solid line]{2004ApJS..154..184S}, UGC5101, a ULIRG with considerable
 9.7$\mu$m absorption \citep[dashed line]{2004ApJS..154..178A}, the prototypical AGN Mrk231
 \citep[dotted line]{2005ApJ...633..706W}, the typical quasar PG1501+106 from
    \citep[triple-dot-dashed line]{2005ApJ...625L..75H}
 and the average mid-IR SED of all starburst galaxies in the IRS GTO
 program from \citep[dot-dashed line]{2006ApJ...653.1129B}.
}
\end{figure}

As noted
in \cite{2005ApJ...634..128T}, the difference in filter bandpasses can
cause substantial differences in the reported flux densities at some
redshifts (up to a factor of 1.5 or even 2) as prominent MIR features
move in and out of the filters.  Figure \ref{fig: iso_ratio}\ shows the
ratio of Spitzer to ISOCAM fluxes for objects with known redshifts (see Section 
\ref{sec: z}).  

The four brightest objects in common between Spitzer and ISOCAM show a larger
offset in photometry than is typical for the sample as a whole.  The
differences may result, in part, from the difference in filter bandpasses as shown
in Figure \ref{fig: iso_ratio}.  All four objects are detected by Chandra
(see Section \ref{sec: xray}), two of them in the hard band.  The sources
do not appear to be variable, however, as all four were also detected by
Spitzer in independent measurements \citep{2005ApJ...634..128T}, and their flux densities agree
within 10\%.

\begin{figure}[t!]
\plotone{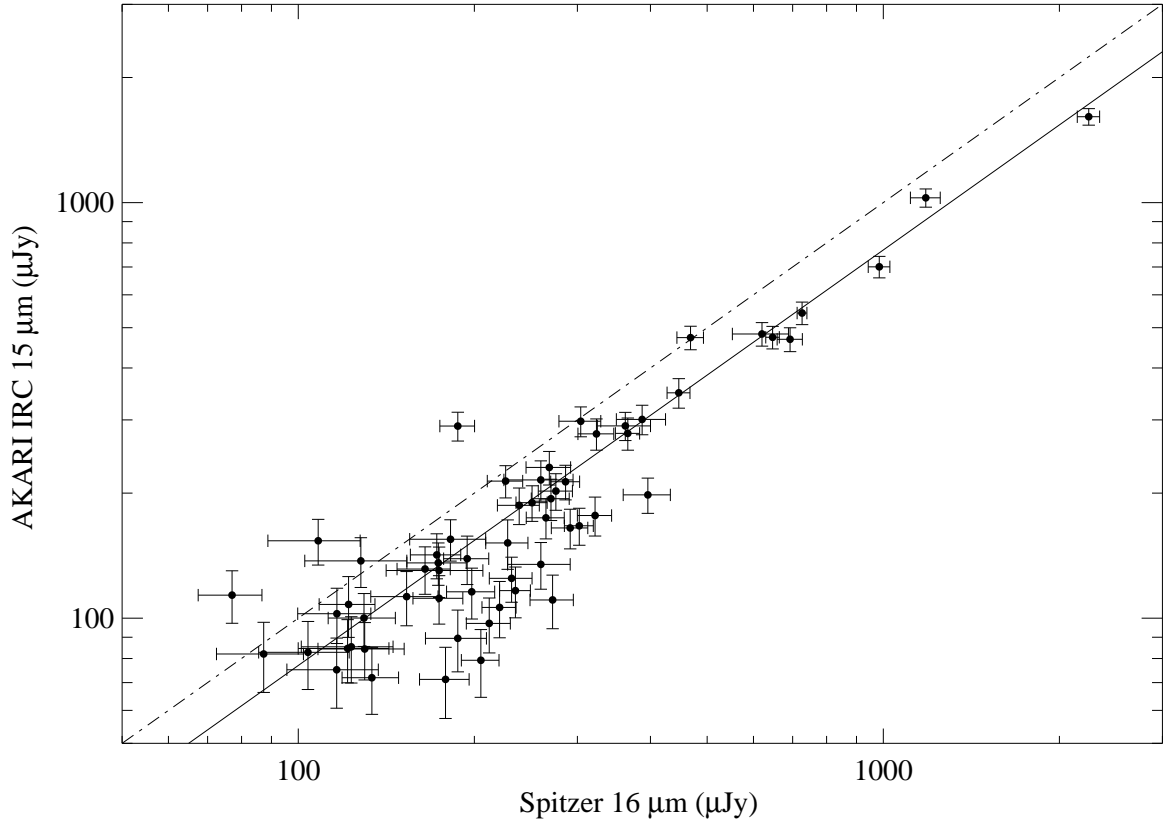}
\caption{\label{fig: compare_akari} The flux density measured with 
{\it Spitzer} PUI in the 16 \mic\ bandpass compared to that measured
with AKARI IRC in the 15 \mic\ bandpass \citep{2009PASJ...61..177B} for objects 
detected by both instruments.  The {\it dot-dashed} line indicates equal flux, and the {\it solid}
line indicates the best-fit offset which corresponds to a factor of 1.3. }
\end{figure}

\cite{2009PASJ...61..177B}\ used the IRC on-board the
AKARI satellite to observe a $\sim
10^{\prime}\times 10^{\prime}$ region partially overlapping with
GOODS-South.  Their catalog contains 67 objects detected at
$>5\sigma$\ in areas which we observed with sufficient coverage (at
least 2 DCEs).  Of these, 60 have associated sources within 3\arcsec\
(two IRC pixels) in the Spitzer catalog.  Figure \ref{fig: compare_akari}\
shows the comparison of Spitzer and AKARI flux densities for matched
sources.  AKARI IRC is also calibrated assuming $\nu f_{\nu} = const$,
but with an effective wavelength of 15.0 \mic, implying a color
correction between the filters of 1.06.  However, the best fit to the photometry
shows a difference in photometry of a factor of 1.3. The filter bandpasses are
slightly different, with the IRC filter being considerably wider,
which may account for some of the difference.

\section{Results}

\label{sec: results}

Extracted sources are given in Tables 1,2 and 3,4 for 
GOODS-N and GOODS-S, respectively.  The tables
include ancillary data from other telescopes, together with
other Spitzer photometry in the IRAC and MIPS 24 \mic\ channels.
Quality flags are given in the tables; caution is recommended in the
interpretation of results based upon sources that do not have quality
1.

Specifically, the tables provide the available data on each source:
position, photometry and uncertainty from Spitzer, HST, and Chandra,
spectroscopic redshifts where available, and quality flags.  The
process for identifying HST, Chandra and redshift associations is
given later in this section.  Details of the columns in the table are
given below:

\begin{itemize}

\item  Column (1) gives the source ID number within the catalog.

\item  Columns (2) and (3) give the right ascension and declination
 (J2000) of
the source.  Positions are reported as the IRAC prior position used as
input to the source extraction.

\item  Column (4) gives the coverage (in number of exposures) of the central pixel
of each source.  Exposure times were 30 (N) and 60 (S) seconds.

\item  Columns (5) through (16) give the IRAC channels 1-4, IRS 16 \mic,
and MIPS 24 \mic\ flux densities and uncertainties for the source, in
units of \mujy.  Uncertainties are given after each photometric point
(e.g. Column 5 is photometry, Column 6 is uncertainty).  The IRAC uncertainties
include a 5\% systematic uncertainty in the calibration (added in quadrature
to the measured uncertainty).  This systematic term dominates the uncertainty
for most objects, especially in channels 1 and 2.  

\item Column (17) gives the spectroscopic redshift for the source,
  where available. Spectroscopic counterparts were usually chosen to
  be the closest optical/NIR object to the 16um position. In a few
  cases, the 16um emission is likely due to a blend of sources, and
  the slightly more distant one is the more likely dominant
  counterpart. A value of -99 indicates that no redshift is available.
  The very few stars with spectra are listed with z=0.

\item Column (18) gives the reference code for the spectroscopic
 redshift (see Section \ref{sec: z}). Redshifts were collated from the
 literature, as well as from observations by the GOODS team (Stern et
 al.\ in prep.).

\item  Column (19) indicates that the source has non-zero flux in at least
one Chandra band (a value of 1 is detected).

\item  Column (20) and (21) give the soft and hard X-ray fluxes, respectively, for the associated {\it Chandra}\ detection.

\item  Column (22) is a star flag, with a value of 1 indicating the
source is likely a star.  A value of 2 indicates that the position of
the IRAC prior is confused between a star and a neighboring galaxy (there
are two such cases in the Northern field).

\item  Column (23) gives the 16 \mic\ concentration index, defined to be the
 ratio of flux within apertures of radii 6 and 2 pixels.

\item  Column (24) gives the number of 8 \mic\ sources within 4 arcseconds.

\item  Column (25) gives the number of 16 \mic\ sources within 5.4 arcseconds.

\item  Column (26) gives the bit-wise quality flag for the source: Bit 0 -- the object was 
determined to be real and included in the catalog; if
only this bit is set, then there are no notes and the object has the best quality; Bit 1 -- possible
confusion because more than one 16
\mic\ source lies within 5.4 arcseconds; Bit 2 -- bad concentration index; Bit 3 --
low SNR but the source was included because the expected SNR was at
least five.  In GOODS-S, we also include a flag for Bit 4, in the case of 
sources with coverage of two exposures instead of three.  Because flags are
assigned bit-wise, sources may have multiple flags set.
So, for example, an object with a flag value of 19 would indicate that Bits 0,1, and 4 had
been set because: (Bit 0) the object was included in the catalog; (Bit 1) the measurement of the object
may suffer from confusion due to a close 16 \mic\ neighbor; and (Bit 4) the object had a coverage of only 
two exposures.

\item  Column (27) flags extended sources which were measured with
aperture photometry instead of the PSF-fit.  A value of 1 indicates
that the source was extended.

\item  Column (28)-(31) give the HST magnitudes for the B,V,I,z bands
 respectively (see Section \ref{sec: hst}).

\item  Column (32)-(35) give the HST uncertainty (in magnitudes) for the B,V,I,z bands respectively.

\item  Column (36) gives the number of I-band sources within 1
 arcsecond radius.

\end{itemize}

We report the detection of 840 (North) and 476 (South) sources.  

\begin{figure}[t!]
\plotone{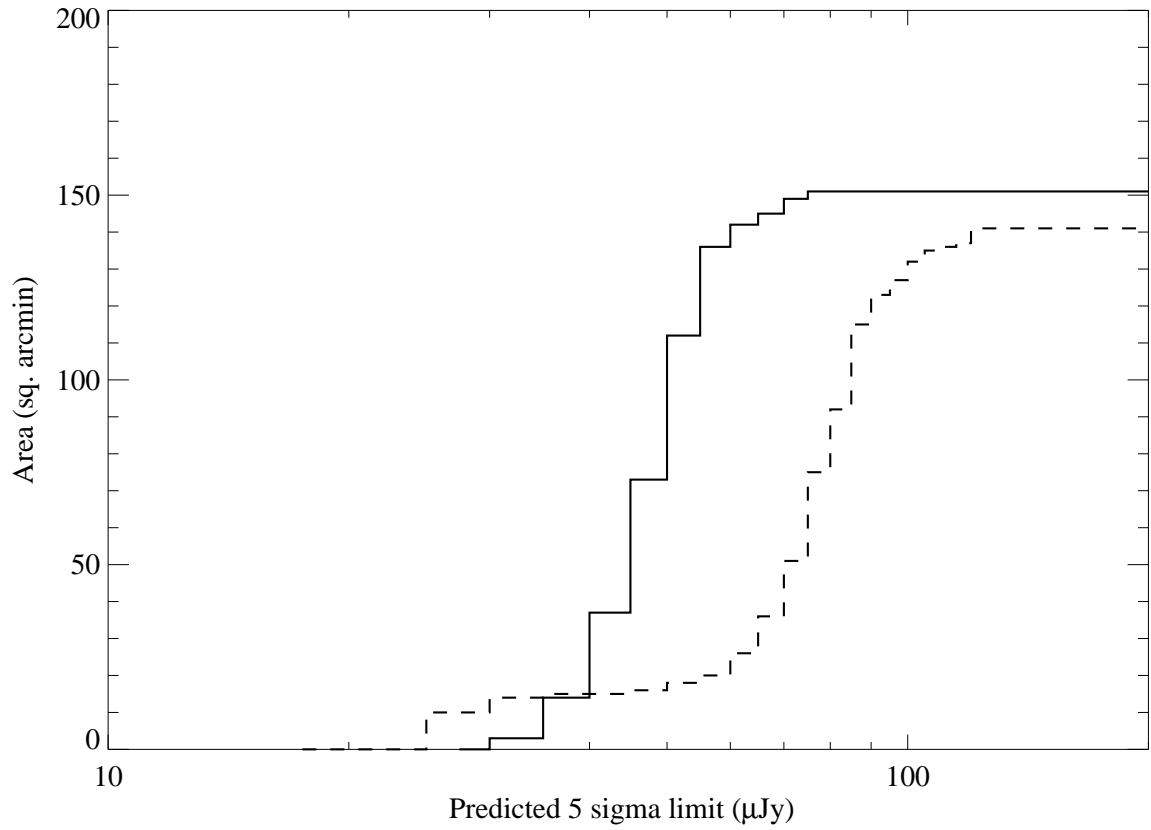}
\caption{\label{fig: depth_area}The area covered at expected sensitivity limits (in $\mu$Jy)
in the North (solid line) and South (dashed line).  Sensitivity predictions are based
upon the SSC exposure time calculator (see text). }
\end{figure}

The depth achieved in the survey is consistent with expectations.  The
5$\sigma$\ limit in the shallow Southern area varies from $\sim 65$\
to 85 $\mu$Jy.  The UDF limit is $\sim 30~\mu$Jy, though the small
area means that there are few objects even in the faintest bin.  The
Northern survey reaches $\sim 40~\mu$Jy.  Figure \ref{fig: depth_area}
shows the area covered in bins of predicted sensitivity.  Table 
\ref{tab: numbers}\ summarizes the number of objects in each field, 
and in some sub-categories (see below).  Table
\ref{tab: depth}\ lists the predicted and achieved depths for the
North, South, and UDF regions.  

In the remainder of this section, we examine the association between 
16 \mic\ sources and other measurements within the GOODS fields.  

\subsection{Optical Photometry}
\label{sec: hst}

We use the optical photometry from the publicly released GOODS version
2.0 catalogs
\citep{2004ApJ...600L..93G}\footnote{http://archive.stsci.edu/prepds/goods/}.
We use the reported {\sc MAG\_AUTO}\ fluxes.  We cross correlate the
positions of 16 \mic\ detections with those of optical sources within
1 arcsecond.  The GOODS catalogs used the $z$-band images for source
detection and then measured photometry in the other three optical
bands. 

Of the 840 sources in the north, 809 have associated optical sources
within 1 arcsecond.  In the South, 465 of 476 sources in the 16 \mic\
catalog have optical associations.  Of the 31 (N) and 11 (S) sources 
without an associated optical detection, 11 (N) and 4 (S) do not have coverage
in the ACS mosaics.  For objects with coverage, 6 (N) and 4 (S) are near the diffraction
spikes from bright stars, and 6 (N) and 1 (S) are in the outskirts of extended galaxies.  
The remaining 8 (N) and 2 (S) sources may be very
red, or they may have a larger uncertainty on the centroid -- either
due to Spitzer pointing uncertainty or to blending with another source
in the IRAC positional priors.
For matched sources,
the possibility of
source confusion is significant when matching MIR and optical sources.
We find 57 (N) and 37 (S) sources have multiple possible optical matches within
1 arcsecond, and we report the closest one to the IRAC position. The
number of close matches is flagged in the table.

\subsection{Redshift Distribution}
\label{sec: z}

Spectroscopic redshifts have been measured for sources in the GOODS
fields by numerous surveys.  The specific
spectroscopic surveys used are given in the tables.

We correlate the positions of detected Spitzer sources with redshift
identifications within about 1 arcsecond radii. In cases for which
multiple optical sources are within 1 arcsecond of the 16 \mic\
target, we usually take the closest positional match. In a few cases
the 16 \mic\ emission is likely due to a blend of sources, and the
slightly more distant one is the more likely dominant counterpart.
Another 21 (N) and 8 (S) objects have optical sources in the
redshift catalogs between 1 and 2 arcseconds away, allowing the
possibility for misidentification in those cases as well. Redshifts
are reported in the table. In the North 701 redshifts are available
out of 826 objects not identified as, or confused with, stars. Most (654) of the
redshifts are for objects with $i_{AB}<24$, and at that magnitude,
redshifts are available for 94\% of 16 \mic\ sources. In the South,
381 redshifts are available out of 466 galaxies detected. Only 36 
of the spectroscopic redshifts in the South are for objects fainter
than $i_{AB}=24$, and 87\% of 16 \mic\ targets down to that magnitude
have redshifts. For comparison, there are 828 redshifts associated
with 5$\sigma$\ 24 \mic\ detections in the North and 814 in the South.

\begin{figure}[t!]
\plottwo{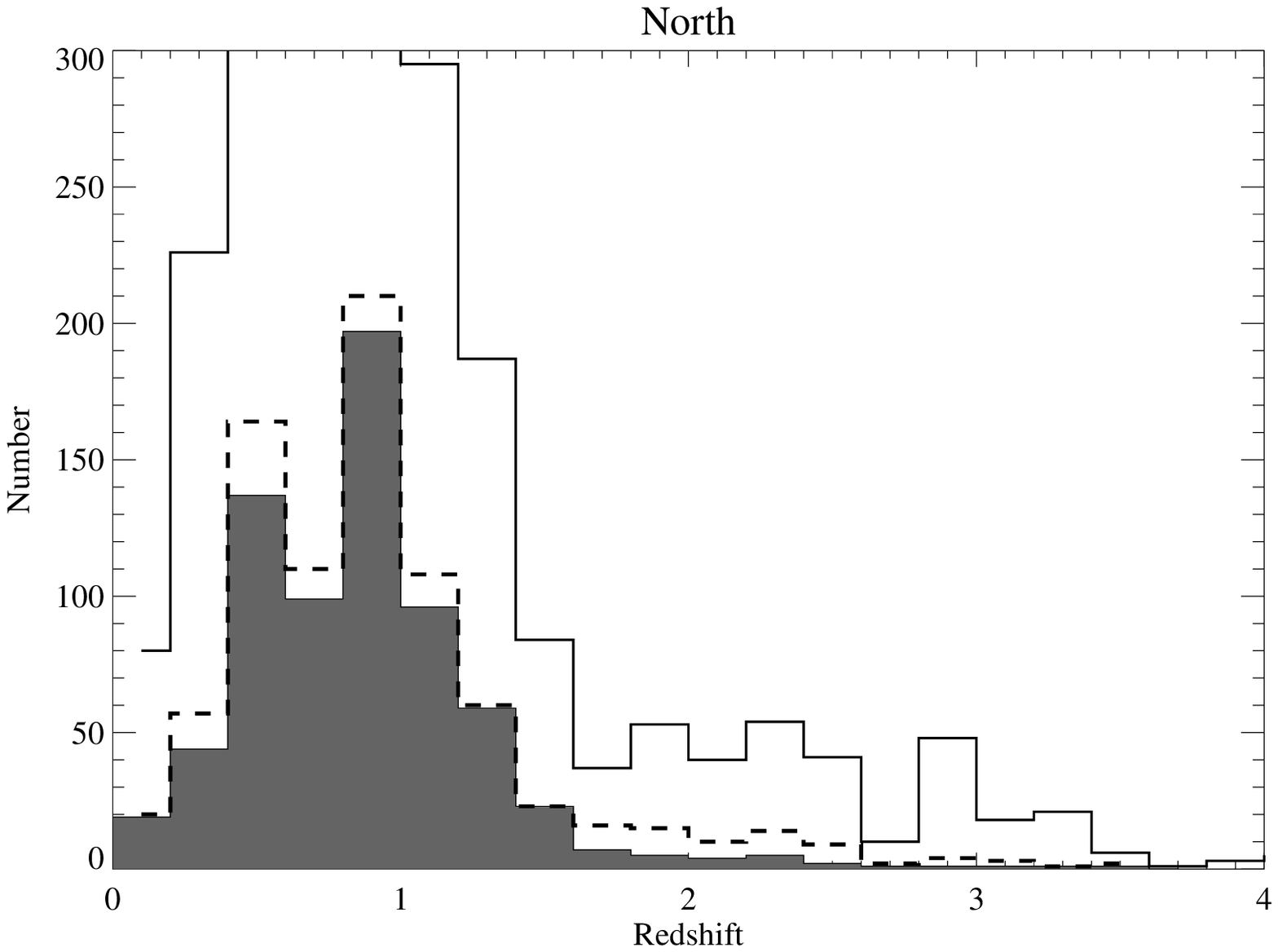}{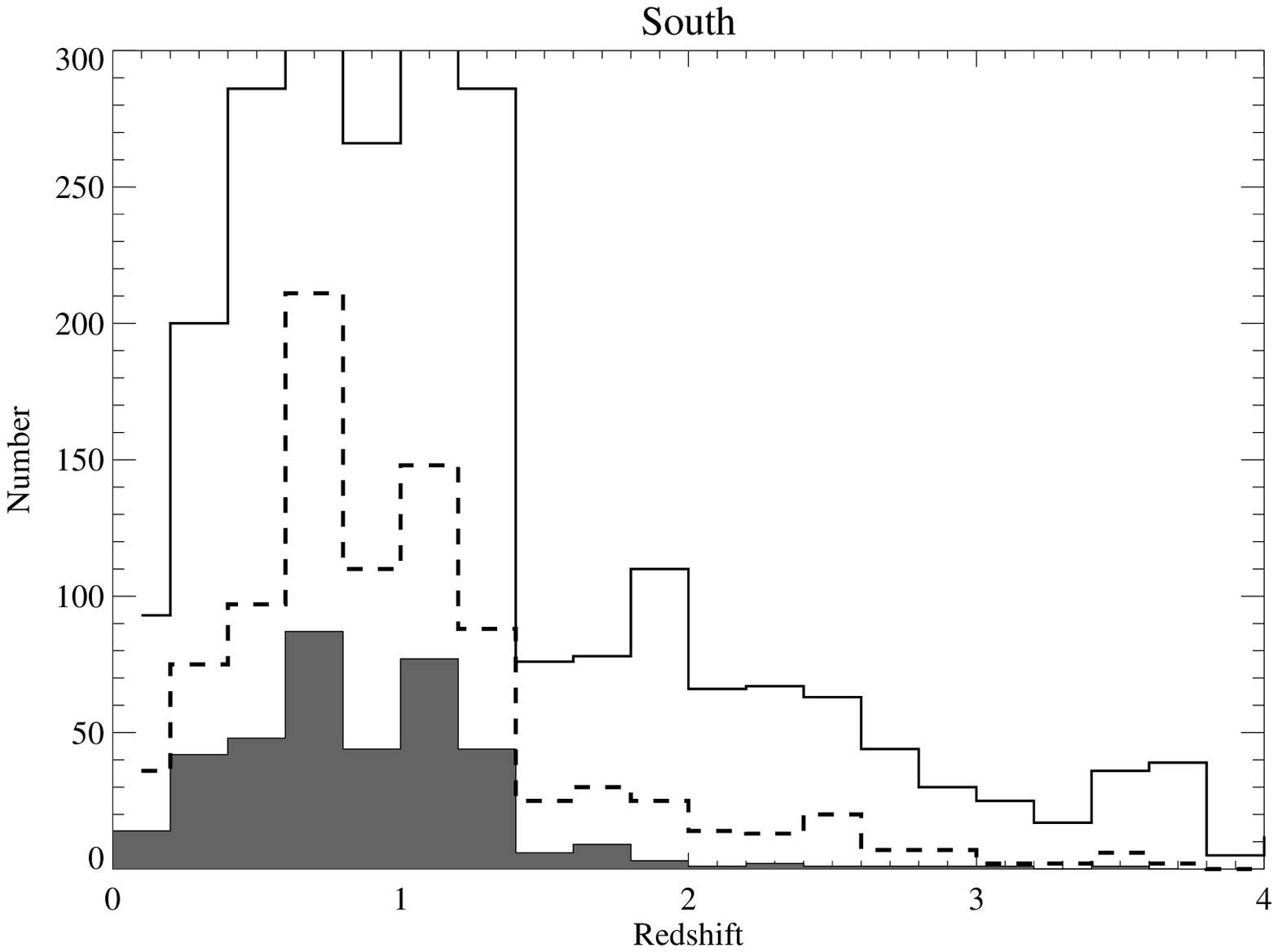}
\caption{\label{fig: zdist} The distribution of spectroscopic redshifts ({\it filled histogram})
for sources associated with 16 \mic\ detections in the North ({\it left}) and South ({\it right}), 
compared to the distribution of sources associated with 24 \mic\ detections ({\it dashed line}) and
the full spectroscopic sample ({\it solid line}; see text) }
\end{figure}

\begin{figure}[t!]
\plotone{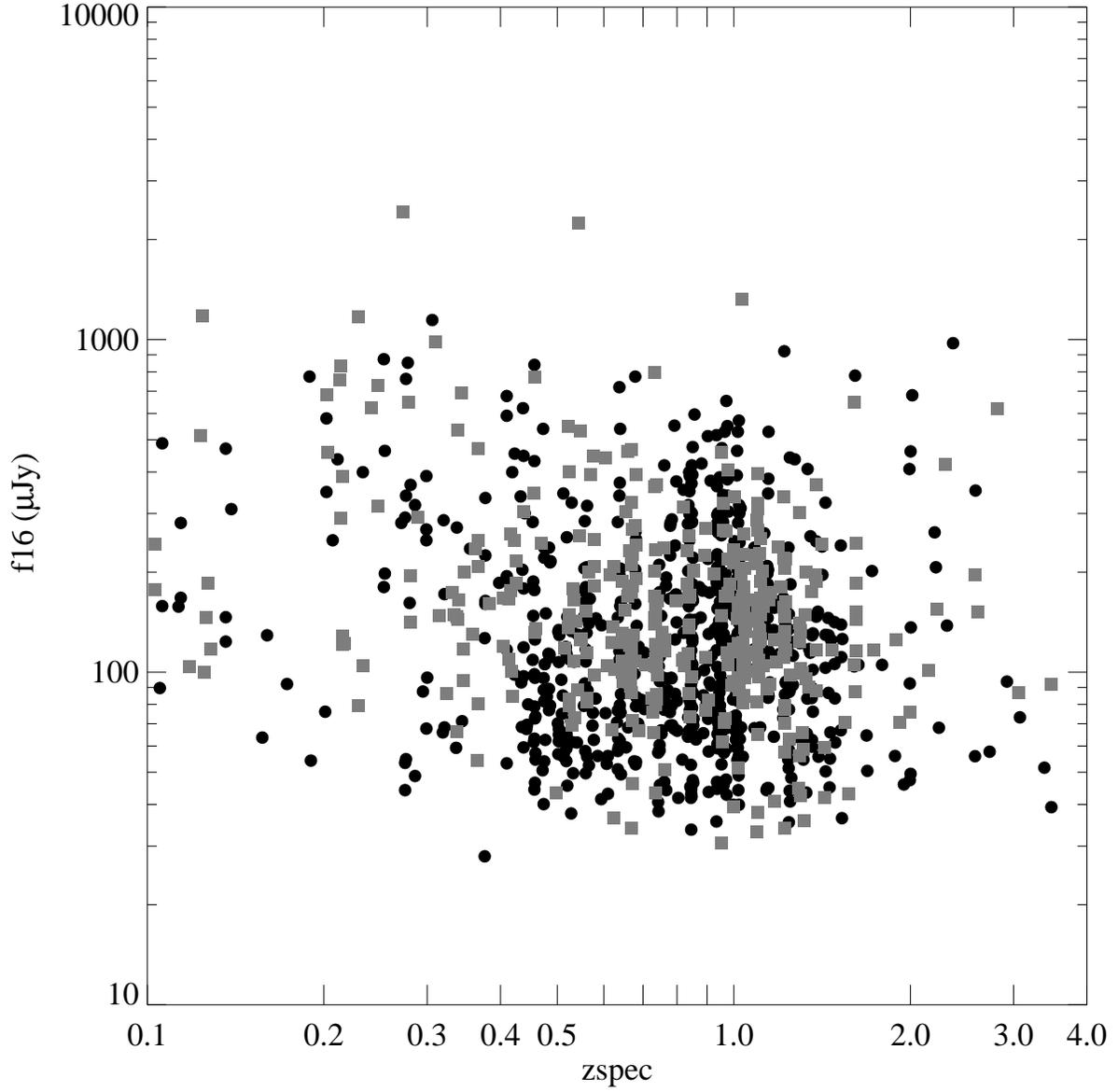}
\caption{\label{fig: f16_z} The flux density of 16 \mic\ detected sources 
versus spectroscopic redshift for GOODS North ({\it black circles}) and 
South ({\it grey squares}).  
}
\end{figure}

Figure \ref{fig: zdist}\ shows the
redshift distribution of detected 16 \mic\ sources.  
It appears that the distribution of associated redshifts is similar
for 24 and 16 \mic\ sources at redshifts less than 1.5.  At higher
redshifts, there are significantly more 24 \mic\ detections, as a
result of both the greater sensitivity of the MIPS observations and
the intrinsic brightening of starburst sources when the strong PAH
features shift into the 24 \mic\ band.  The similarity of the
distributions at lower redshifts suggests that the distribution may be
a stronger function of the optical limits on obtaining redshifts
rather than the MIR observations themselves.
Figure \ref{fig: f16_z}\ shows the 16 \mic\ flux density versus redshift.

The median spectroscopic redshift, excluding stars, for 16 \mic\ sources
is 0.85 (N) and 0.82 (S), and the mean is 0.86 in each field. About 30\% of 
redshifts in the North are at $z>1$, and about 35\% in the South. About
2\% of the sources in each field (13 North, 6 South) have $z_{spec}>2.0$.  Most of these
($\sim$half in the North and 5/6 in the South) are identified as AGN, as
described below.  The highest spectroscopic redshifts for 16 \mic\
sources are 3.48 (N) and 3.47 (S).

\subsection{X-ray Sources}
\label{sec: xray}

The {\it Chandra}\ 2 Msec surveys of the GOODS North and South
fields, respectively, are the deepest X-ray observations taken to
date. We compare the 16 \mic\ survey to the {\it Chandra}\ catalogs of
\cite{2003AJ....126..539A} and \cite{2008ApJS..179...19L}. There are
308 (N) and 293 (S) X-ray
sources within the GOODS area surveyed at 16 \mic. Of these, 117 (N)
and 92 (S) are associated with 16 \mic\ sources, comprising 14\% and
20\% of the 16 \mic\ samples. Sources that are detected in the {\it
 Chandra}\ catalogs are flagged in the table.

\section{Discussion}

\subsection{Active Galactic Nuclei}
\label{sec: agn}

The detection of sources at 16 \mic\ selects both strongly
star-forming galaxies and those which host AGN.  Most sources in the
present survey lie at redshifts below 2, which ensures that the PUI
passband samples the wavelength range that covers emission by dust
rather than direct stellar light.  At $z>1$, the PUI band is
increasingly dominated by hot dust characteristic of AGN, and at $z>2$
the sensitivity of the survey is mostly limited to such sources.

Hard X-rays are usually indicative of an AGN.  The ultradeep {\it
 Chandra}\ data select many of the AGN within GOODS. There are 96 (N)
and 58 (S) PUI sources associated with detections in the hard band.
Given the depth of the CDFs, a few of these objects 
may have X-rays from purely star-forming galaxies. For example,
\cite{2006ApJ...640..603T} excluded sources with $L_X<10^{42}$\ ergs
s$^{-1}$\ when selecting AGN in GOODS. The fraction of such sources is
small amongst 16 \mic\ targets, and the exact nature of those sources
is uncertain, so we retain them as possible AGN in this analysis.

Some galaxies hosting AGN, even those whose bolometric luminosity is
dominated by them, are undetected even by {\it Chandra}.
These sources may be selectable using their MIR color, when AGN heated dust
dominates over stellar light.  In many cases,
these will be have power-law SEDs \citep[e.g.][]{2007ApJ...660..167D}.
We perform a $\chi^2$\ fit to the IRAC$+16~\mu$m colors to identify power-law
sources.  We fit all SEDs with a power-law, $\nu^{-\alpha}$, and select those
with $\alpha>0.25$.  We do not set a limit on the goodness-of-fit, but we exclude
sources where the channel 3 or 4 flux does not exceed the channel 1 flux. Using 
only IRAC photometry with 16 \mic\ does not significantly change the results. We find 
about 10\% more AGN when combining X-ray with power-law selection than when
using x-ray selection alone.  Conversely, we find that about 45\% of hard X-ray
sources have power-law SEDs.

A
similar fraction was measured by \cite{2006ApJ...642..126B}, who found
that $\sim 40$\% of X-ray sources in the extended Groth Strip
have red power-law SEDs in the IRAC bands.

\begin{figure}[t!]
\plotone{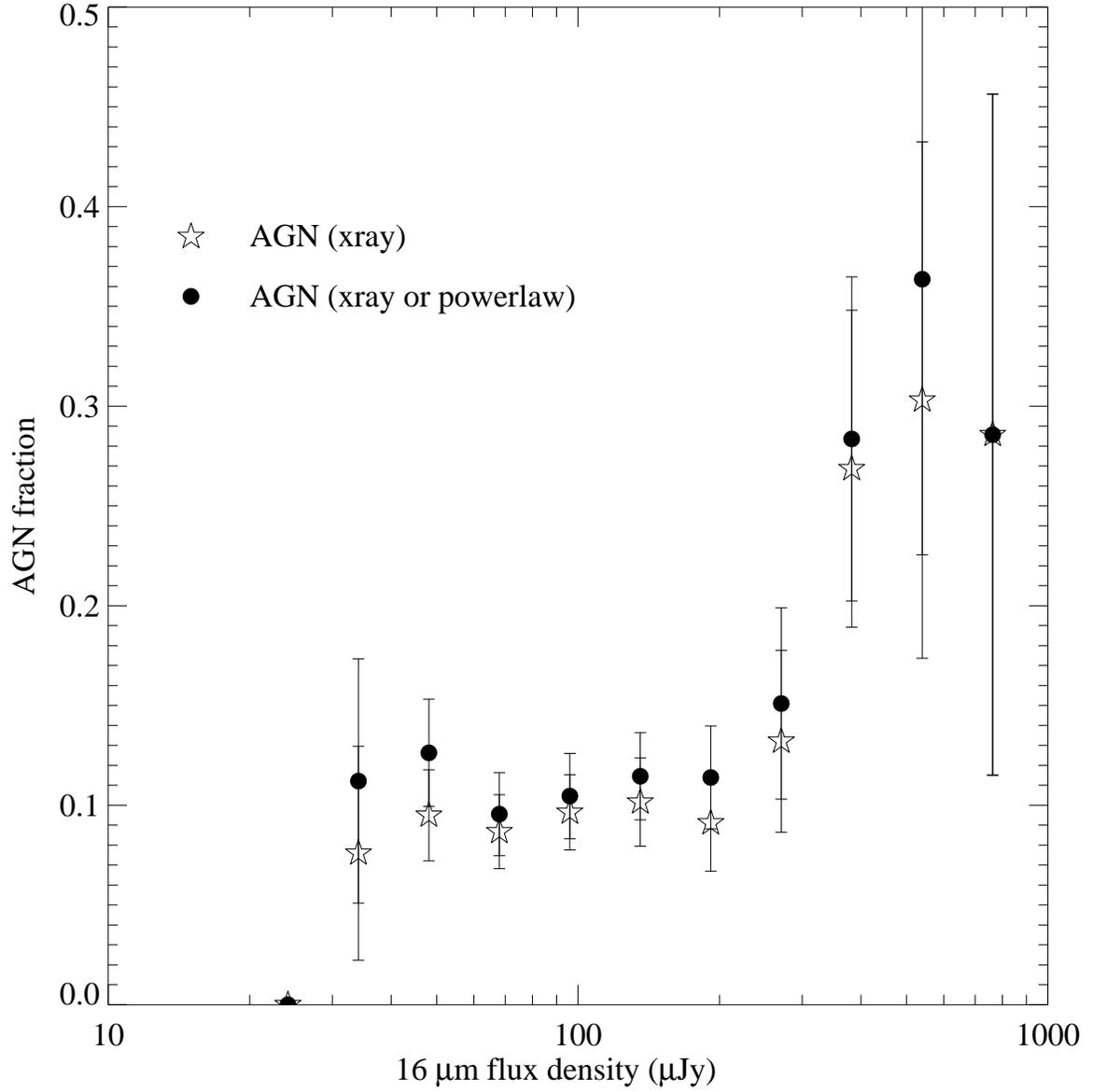}
\caption{\label{fig: agnfrac} The fraction of 16 \mic\ sources
 identified as AGN ({\it filled circles}) by the detection of hard
 X-rays ({\it open star symbols}) and/or power law slope.}
\end{figure}

Combining the two selections, we find 105 (N) and 65 (S) AGN. These
numbers are likely to be a slight underestimate, as they may not
include Type 2 AGN whose IRAC colors are not true power-laws.. Nonetheless, we find that
about 15\% of 16 \mic\ sources are galaxies hosting AGN. Figure
\ref{fig: agnfrac}\ shows the fraction of sources with AGN in bins of
16 \mic\ flux density. The fraction is higher for the brighter
sources. 

Amongst rare, bright sources, obscured AGN may be more common.
\cite{2005ApJ...631..163S}\ identify QSOs and Seyfert 1 galaxies as
objects with red IRAC colors, but their selection is contaminated by
star forming galaxies at the faint fluxes, such as those in the GOODS
survey. \cite{2007ApJ...660..167D}\ find that when using the IRAC
color-color selection only 55\% of MIR power-law AGN are detected in
the X-ray.

\cite{2006ApJ...640..603T} find a decreasing AGN fraction with
decreasing 24 \mic\ flux for sources in the GOODS fields, down to
$\sim8$\% at $<100$\ \mujy\ using purely hard X-ray selection. The
fraction of X-ray sources at 16 \mic\ is similar, as expected given
the high rate of detection of {\it Chandra}\ sources.
\cite{2008ApJ...687..111D}\ also find a decreasing AGN fraction with
decreasing 24 \mic\ flux, down to about 10\% X-ray AGN. They then
expand the estimate to MIR-selected AGN and estimate about 15--25\% at
100-300 \mujy, similar to our estimate given the large error bars.

Table \ref{tab: numbers}\ summarizes the numbers of AGN per field.

\subsection{16 to 24 \mic\ Color}

One of the most prominent MIR spectral features is the broad (full
width $\sim 2$\ \mic) silicate absorption trough at 9.7 \mic.
Attenuation at this wavelength can approach an order of magnitude in
typical ultraluminous IR galaxies \citep[e.g.][]{2004ApJS..154..178A}. As a result, this
feature will significantly depress the photometry measured in a
broad-band filter. \cite{2005MNRAS.357..165T}\ suggested that the 16
to 24 \mic\ ratio can be used to identify ``silicate-break'' galaxies,
at redshifts $1.1<z<1.6$ where the silicate absorption is solidly
within the MIPS 24 \mic\ bandpass.  

\begin{figure}[t!] 
\plotone{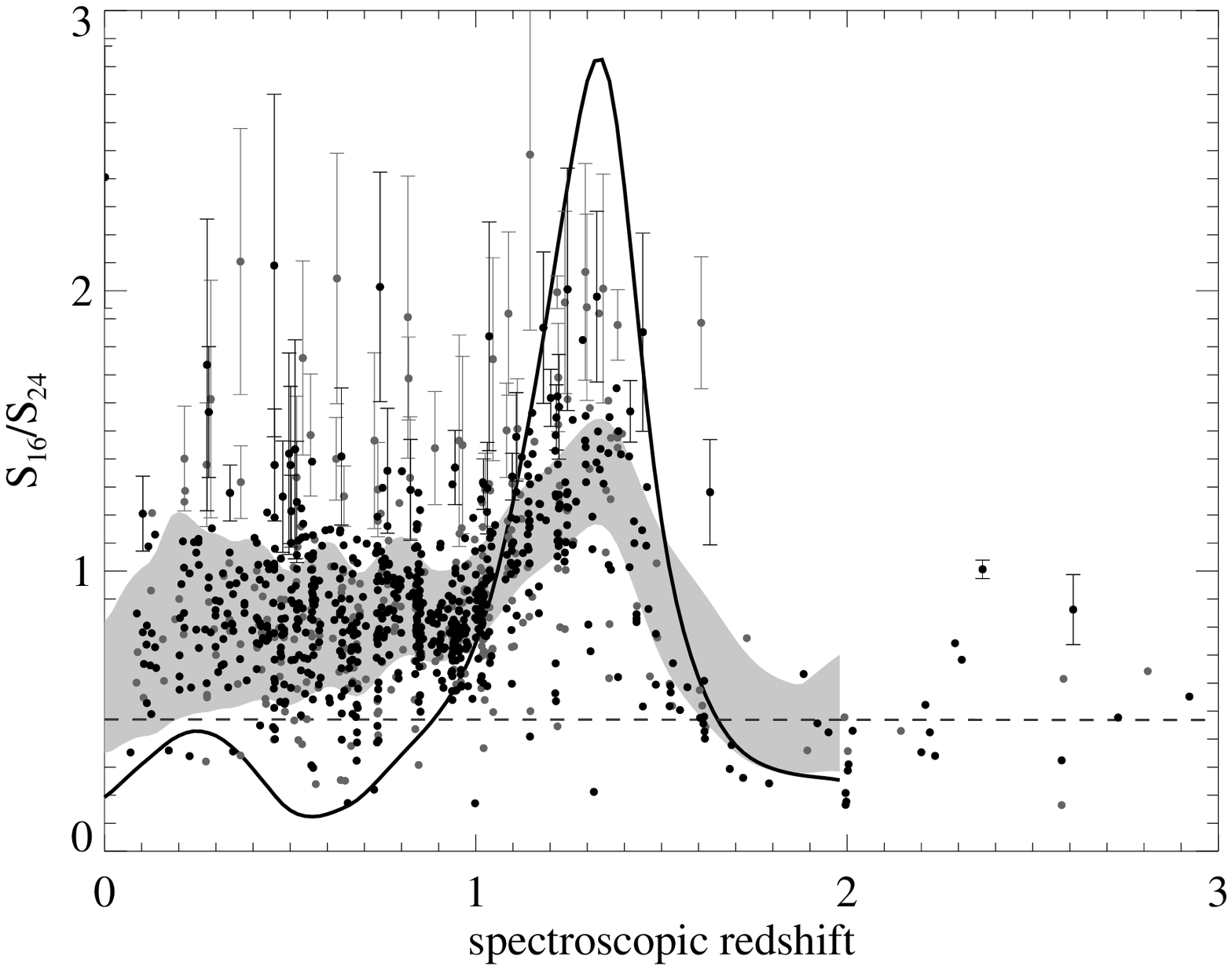} 
\caption{\label{fig: ratio} The ratio of 16 to 24 \mic\ flux
  densities as a function of spectroscopic redshift for high quality
  (flag=1; {\it black points}) and all sources ({\it grey points}).
The shaded 
  region indicates the range of values expected from the starburst
  templates of \cite{2006ApJ...653.1129B}\ and \cite{2007ApJ...656..770S}.  The
  ratio for Arp220 \citep[using the spectrum
  from][]{2007ApJ...656..148A} is plotted ({\it solid line}), as
  is the value for a power law spectrum with $f_{\nu}\sim \nu^{-2}$\
  ({\it dashed line}).  Uncertainties are plotted for objects
  with colors more than $1\sigma$\ bluer than expected from the local
  starburst templates.}
\end{figure}

Figure \ref{fig: ratio}\ shows the $S_{16}/S_{24}$\ ratio of catalog
sources for which spectroscopic redshifts are available, and which are
solidly detected in the 24 \mic\ band.  We apply a
small color correction (4\%) to the MIPS 24 \mic\ flux densities in
order to account for the difference in the way the instruments are
calibrated.  The PUI is calibrated assuming a reference spectrum of
$\nu F_{\nu}=const$, but MIPS is calibrated using a $10^4$\ K black
body (see the MIPS Instrument
Handbook).

Much of the scatter in the flux ratio is expected due to variation in
the source SEDs.  We demonstrate this by calculating the expected
ratio for redshifted local template sources.
\cite{2007ApJ...656..770S}\ obtained Spitzer IRS spectra for a range
of local starbursts, and found a large variation in both PAH
emission strength and silicate absorption depth.  They find that the
relative strength of features can vary by a factor of two in star
forming galaxies ($L_{IR}<3\times10^{11}~L_{\odot}$)\ depending upon
source properties, though they note the variation should be somewhat
less in more luminous sources.  \cite{2006ApJ...653.1129B}\ also
measured local starbursts, notably NGC 7714, and calculated an average
starburst spectrum, which is redder than the Smith et al.\ templates,
and so results in a lower expected ratio.  In the plot, we indicate
the expected ratio for these templates as a shaded region. Similar
results have already been reported by \cite{2009PASJ...61..177B}
using the AKARI measurements in GOODS-S.  They find that local
starburst templates mostly reproduce the flux ratio well, but that
there are some discrepancies at higher luminosities.  The AKARI survey
was limited to $z<1.2$.

\cite{2007ApJ...656..148A} calculated the expected 16 to 24 \mic\
ratio for local ultraluminous IR galaxies (ULIRGs), and found that it is significantly lower than 
for starbursts at
most redshifts because ULIRGs have a red continuum.  However,
in the redshift range for which the silicate absorption falls in the 24 \mic\
passband, the ratio can be extreme ($>3$\ for some ULIRGs).  In the
Figure we plot the ratio using Arp 220 as a representative ULIRG
template.  Finally, we note that AGN lacking strong silicate or PAH
features will have red colors at all redshifts.  Figure \ref{fig: ratio}\ also
shows the ratio for an AGN with a very red ($\alpha=-2$) power law
SED ($f_{\nu} \propto \nu^{\alpha}$).

For clarity, error bars are omitted from points in the figure within
the expected region.  Typical uncertainties are 0.1-0.2 in the flux
ratio.  Instead, we only plot individual uncertainties for objects
with 16/24 colors more than $1\sigma$\ bluer than the expected
regions.  There is significant scatter in the plot, but a trend at
$z>1$\ indicating the presence of silicate absorption is present.

A number of sources at $z<1.1$\ are anomalously blue, though most of them
are likely the result of measurement error (large uncertainties) or data quality
issues (most have quality 
flag warnings, as shown in the figure).  The rest may have
SEDs that are bluer than local starbursts because the 24 \mic\ band is tracing
longward of 12 \mic\ emission and their warm dust contribution is low. 
 
Objects at $1.1<z<1.6$\ with a ratio bluer than $S_{16}/S_{24}=1.2$\
lie solidly in the region expected for silicate absorption.
\cite{2005ApJ...634L...1K}\ used the same selection when identifying
silicate absorbing sources in early Spitzer observations of the NOAO
Deep Wide Survey field. However, many objects at lower redshifts fall
within the same cut. Selecting objects with a ratio $>1.4$\ eliminates
many, though not all, of the low redshift interlopers. If we consider
only sources with a 16 \mic\ quality flag of 1 and an available
spectroscopic redshift, then 10 of 44 selected sources are outside the
redshift range for silicate absorption. For comparison, $\sim$80\% of
16 \mic\ sources with spectroscopic redshifts are at $z<1.1$. If we consider all sources
which are securely detected at 24 \mic\ and have available redshifts (regardless of quality flag),
then 33 of 91 sources are possible interlopers; however, about half of these
have quality flag with either Bit 1 or Bit 2 set, indicating possible confusion or a
bad concentration index.  We note that not all objects with blue
ratios are true silicate absorbers.  Many, perhaps most, are
star-forming galaxies where the blue color results from the dip
between the 7.7/8.6 and 11.2/12.7 \mic\ PAH complexes \citep{2007ApJ...656..770S}. 

A few objects lie at redshifts higher
than the range where silicate absorption falls in the 24 \mic\ band but still 
have colors bluer than expected for either AGN or the starburst template.
These objects could have buried AGN causing large hot dust emission at 5-6 \mic, or they
could have anomalously strong 6.2 PAH features.

\begin{figure}[t!]  
\plotone{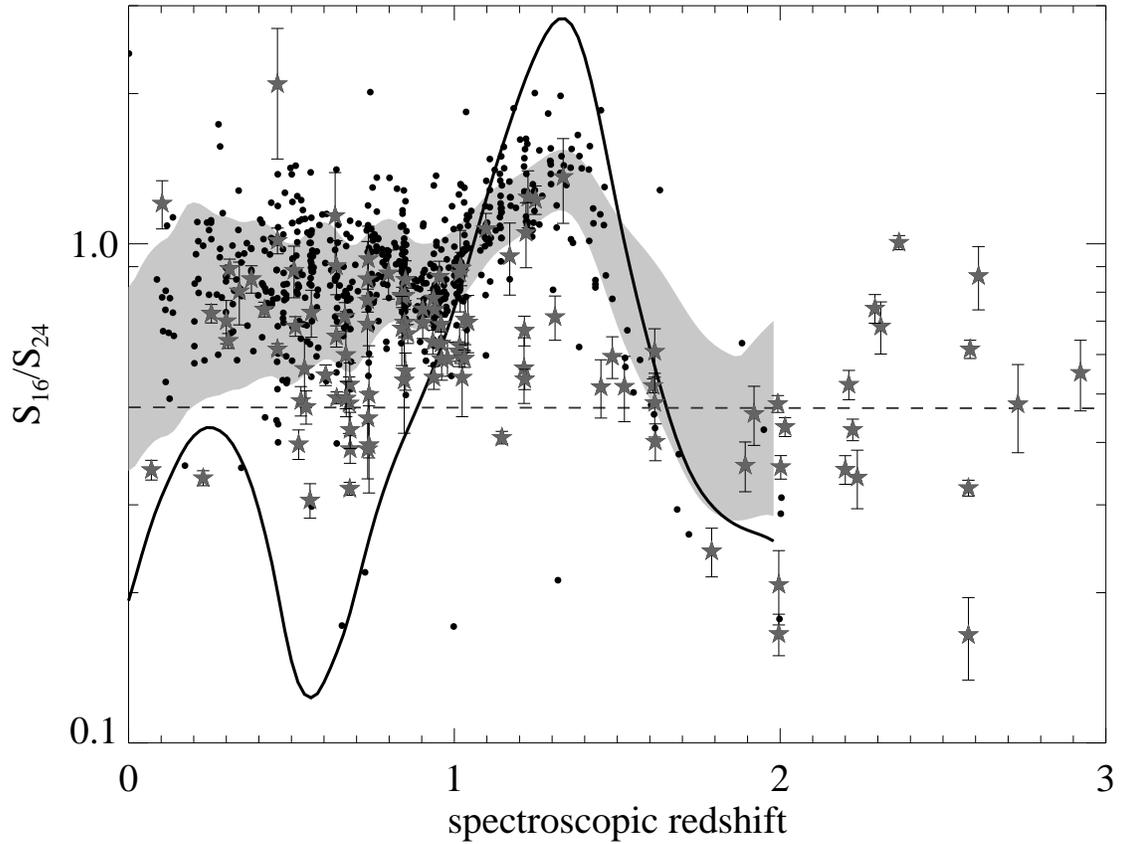} 
\caption{\label{fig: ratio_agn} As in Figure \ref{fig: ratio}, we show the 
flux ratio for high quality sources (flag=1; {\it black points}) and highlight
objects identified as possible AGN through hard X-ray detection or
power
law slope
 ({\it star-shaped symbols}, see section
 \ref{sec: agn}).}
\end{figure}

Figure
\ref{fig: ratio_agn}\ shows the flux ratio for objects selected as AGN
by either power-law SEDs or hard X-ray detection.  The $S_{16}/S_{24}>1.4$\
selection does not appear to select objects that are dominated by known AGN.  
In the Figure, we also 
see a number of AGN sources with very low $S_{16}/S_{24}$\ ratios.  At $z\sim 0.6$, some of
these may be explained by silicate absorption in the 16 \mic\ band.  Of course,
the ratio will be low for objects with red SED as well, so the silicate selection is
not as clean.

\begin{figure}[t!]  
\plottwo{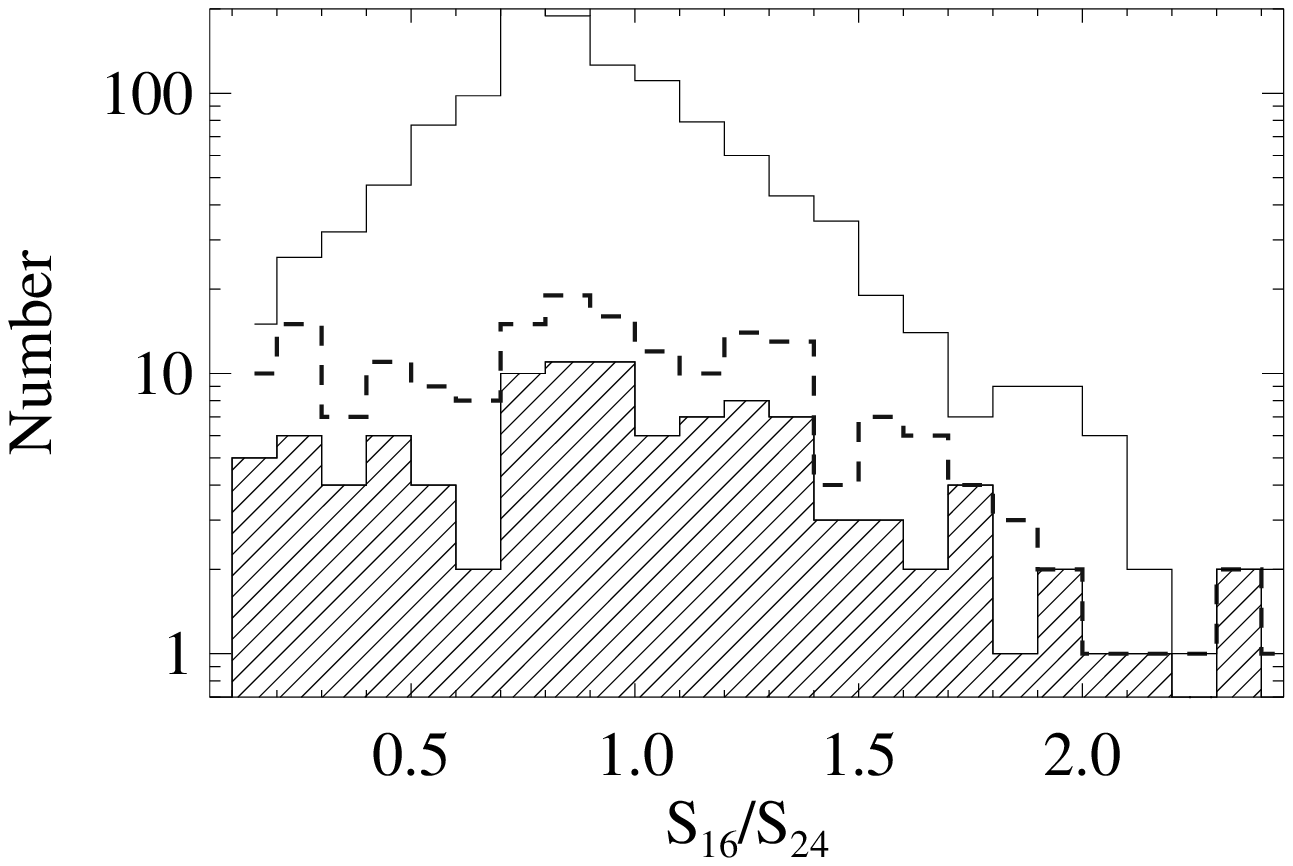}{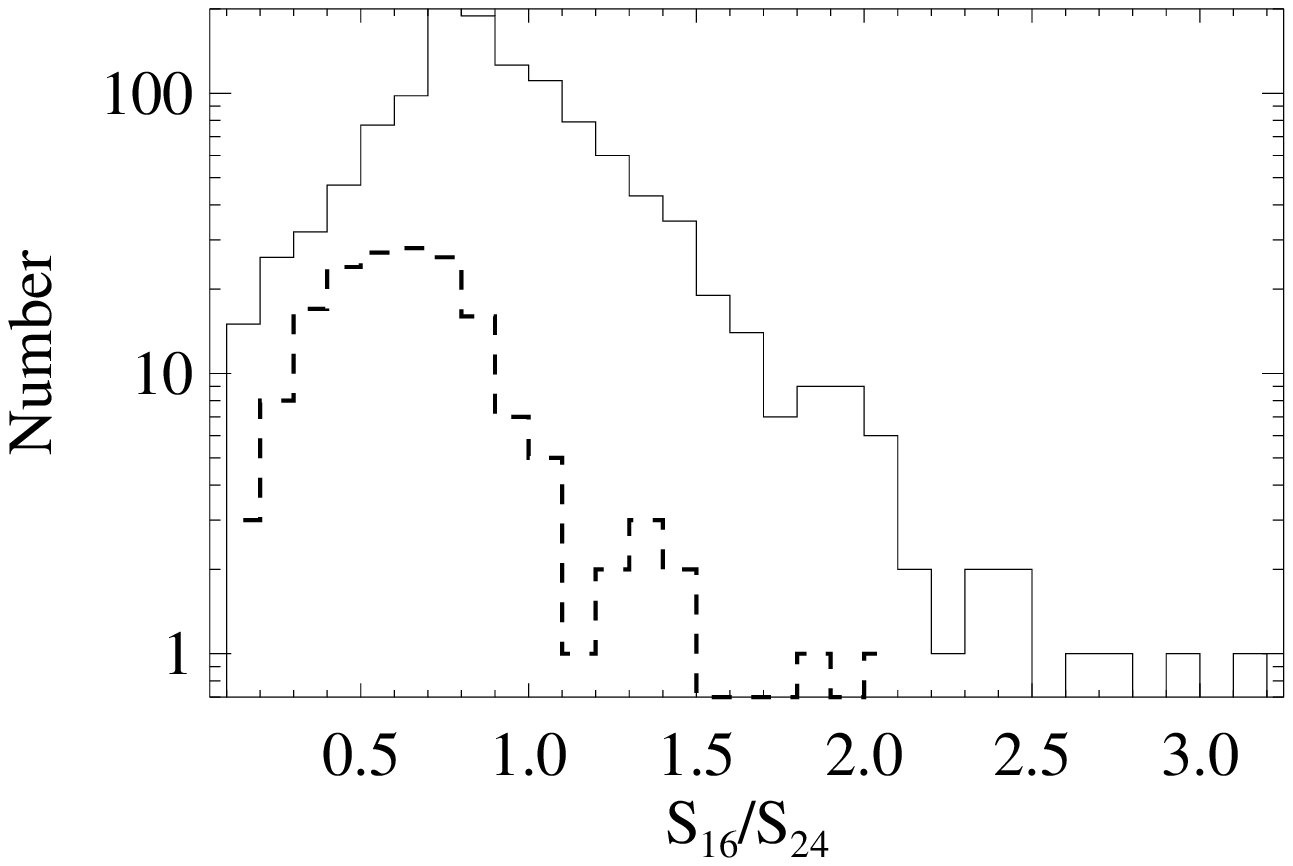} 
\caption{\label{fig: ratio_hist} Histogram of 16 to 24 \mic\ flux
 ratios for all objects ({\it solid line}) and objects without
 spectroscopic redshift (left; {\it filled histogram} for objects with
 photometry quality
flag equal to 1; {\it dashed line} for all objects without redshift) and AGN (right; {\it dashed line}).}
\end{figure}

Figure \ref{fig: ratio_hist}\ shows histograms of the flux ratio
for the 16 \mic\ sample, and the AGN subset. Applying the ratio cut at
1.4 (and excluding sources with other spectroscopic redshifts), we
find 107 blue sources, of which possible 68 have a quality flag of 1, which could
indicate that they are $z\sim 1.3$\ sources.
Table \ref{tab: numbers}\ compares these numbers with the AGN selected for each field.

\subsection{Number Counts}

We calculate galaxy number counts for each field.  First, we determine
the area, $A_i$, over which each source could have been detected
assuming our estimate of the sensitivity based upon the depth of
coverage.  Then, we sum the reciprocal of the area for each source
within logarithmically spaced bins of flux and divide by the binwidth,
$\delta_f$.

\begin{equation}
nc = \frac{1}{(\delta f)_{bin}} \sum_{i_{bin}} \frac{1}{A_i}
\end{equation}

This calculation does not include the effects of source confusion 
which could lead to slightly under-counting the number of galaxies.
We estimate the uncertainty on the counts for each bin as the Poisson
error in the measurement of the number of sources in the bin. 

Completeness corrections were calculated using a Monte-Carlo
simulation, following \cite{2004ApJS..154...80C}.  Separate
simulations were performed for the three depth tiers of the survey --
GOODS-North, GOODS-South, and the UDF.  Artificial sources were added
to the original data images and recovered.  These sources were placed
at random positions.  The flux distribution of simulated sources was
flat in $log(f_{\nu})$.  The fluxes varied from 20 to 1000 $\mu$Jy.
To avoid confusion of simulated sources, only a small number were
added at a time and the simulation was repeated many times.  In the
North and South, 50 objects were added to the images at a time, with
the simulation repeated to build up ten thousand simulated input
sources.  In the UDF, only 15 sources were added at a time.  Sources
were recovered using the same positional prior code used for the
catalog generation, discarding sources below $5\sigma$.  The
completeness of recovered sources is seen to be worst at the faint
fluxes, as expected.  There is considerable incompleteness at
relatively bright fluxes, as well, due to confusion of simulated and
real sources.  This effect appears to be on the order of 5--10\%.

A matrix $P_{ij}$\ for the output flux distribution of the artificial
sources was generated, where $i$\ is the input flux and $j$ is the
recovered flux \citep{1995ApJ...449L.105S}.  The classical
completeness for sources in the $i^{th}$\ bin is the ratio of number of
recovered sources in that bin to the sum over all $j$\ for that bin. 
The observed catalog of sources in the real image was then distributed
among the flux bins.  The $P_{ij}$\ matrix was renormalized such that
the sum over $i$\ for each $j$ was equal to the number of detected
sources in that flux bin.  The completeness corrected counts in each
flux bin $i$\ is then the sum over $j$\ of the renormalized $P_{ij}$\
matrix.

In the North, we estimate completeness to be $\sim 50$\%
at 40 \mujy\ (the approximate limit of reported sources), and 80\% at
60 \mujy.  In the South, we estimate 80\% completeness at 80 \mujy\ in
the shallow survey, with a steep fall off at fainter fluxes due to
poor coverage.  In the calculation of the number counts, we do not use
the shallow Southern survey below 65 \mujy.  In the UDF, we estimate
50\% completeness at 30 \mujy\ (the limit of reported sources) and 80\%
at 50 \mujy.  

\begin{figure}[t!]
\plotone{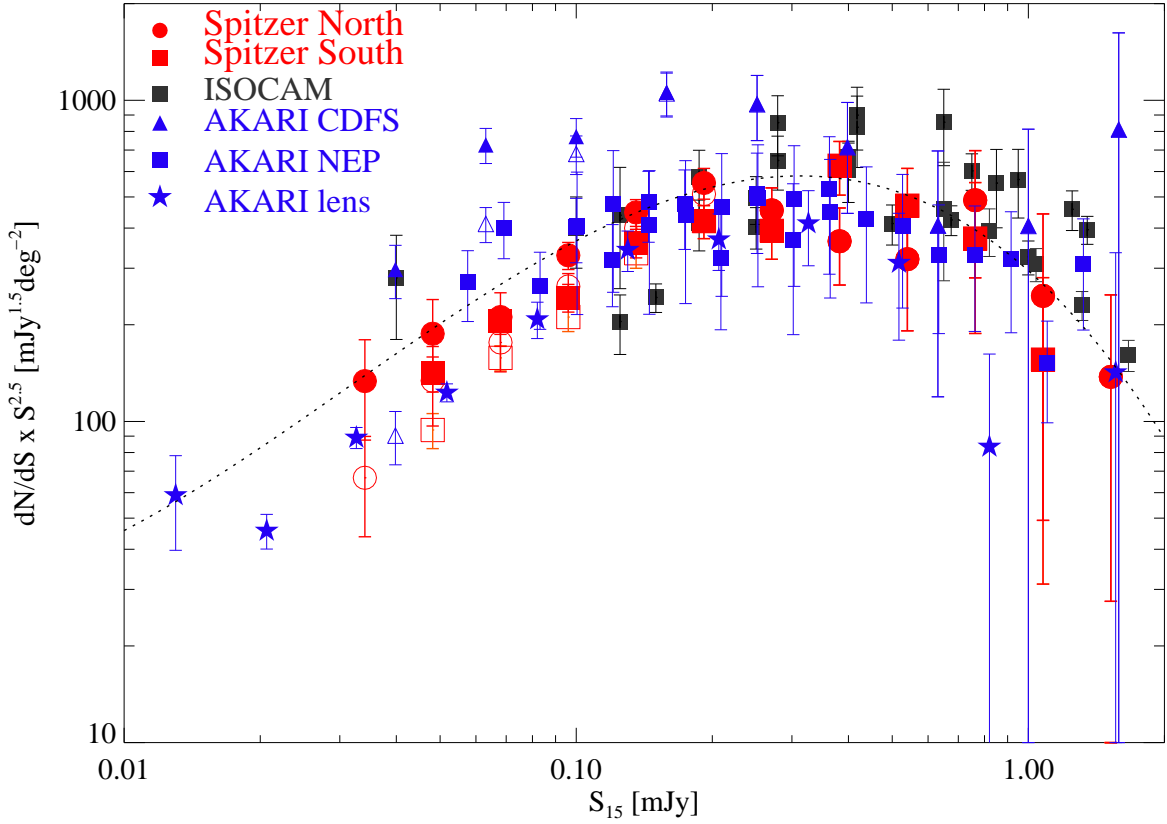}
\caption{\label{fig: diffcounts} Differential 16 \mic\ number counts
 measured by Spitzer ({\it filled red symbols}\ have completeness correction applied, 
{\it open red symbols} do not), AKARI ({\it blue symbols}), and ISO
({\it dark grey
 squares}). The Euclidean slope has been removed. Poissonian error
 bars are shown.  ISOCAM points include HDF-N, HDF-S, and the Marano
 surveys \citep[][and the references therein]{1999A&A...351L..37E}; the
 gravitational lensing cluster survey \citep{1999A&A...343L..65A}; the
 European Large-Area ISO Survey \citep[ELIAS-S1;][]{2002MNRAS.335..831G};
 and the Lockman Deep and Lockman Shallow surveys
 \citep{2004A&A...427..773R}. 
We plot AKARI points from the completeness-corrected counts measured in the North ecliptic pole 
\citep[{\it filled
  squares};][]{2007PASJ...59S.515W,2010A&A...514A...8P}.  The \cite{2009PASJ...61..177B}\ survey of 
GOODS-South ({\it filled blue triangles}\ are counts with completeness
correction, {\it open blue triangles}\ 
are the counts without) and we exclude bins with less than 20\%
completeness.  We also show the completeness-corrected AKARI counts with lensed
objects in Abell 2218 \citep{2010ApJ...716L..45H}.  We also show the fit to the 
combined Spitzer 16 \mic\ counts from the two GOODS fields ({\it dotted line}; see Section \ref{sec: cirb}) }

\end{figure}

Figure
\ref{fig: diffcounts}\ shows the measured number counts, including the
completeness correction described below.  The figure also compares 15
\mic\ number counts from ISOCAM and AKARI, but without
color-correction.  \cite{2009arXiv0901.3783L} showed that these number
counts are dominated by low redshift objects ($z<0.5$) at fluxes above
200 \mujy, and by moderate redshift ones ($0.5<z<1.5$) at fainter
levels; they infer a small contribution ($<20$\%) from higher redshift
sources in bins below 100 \mujy.

In the figure, we see a significant difference between the
counts in the Northern and Southern fields.  The
counts peak around 0.4 mJy in the North, but around 0.2 mJy
in the South.  It is likely that this effect is the result of
cosmic variance.  However, the distribution of redshifts (see Section \ref{sec: z})
shows no clear evidence for 
an over-density in the Northern field to explain the difference.  \cite{2009ApJ...703..222L}\ 
noted a similar effect in comparing GOODS-North counts at 24 \mic\ to those in the
(wider area) COSMOS field.  

\cite{1999A&A...351L..37E} measured a faint-end slope of $\alpha=-1.6$\ from
the ISOCAM 15 \mic\ counts. They defined the faint-end to be bins with $S_{15}<0.4$\ mJy. 
In the same range, we measure a slightly steeper slope of -1.9 when fitting the
counts from both the North and South.  However, the fit is dominated by the high 
significance bins between 0.1 and 0.4 mJy, where the two fields have different peaks.
If we consider only $S_{16}<0.2$\ mJy, then $\alpha=-1.7$\ is consistent with the data.

\begin{figure}[t!]  
\plotone{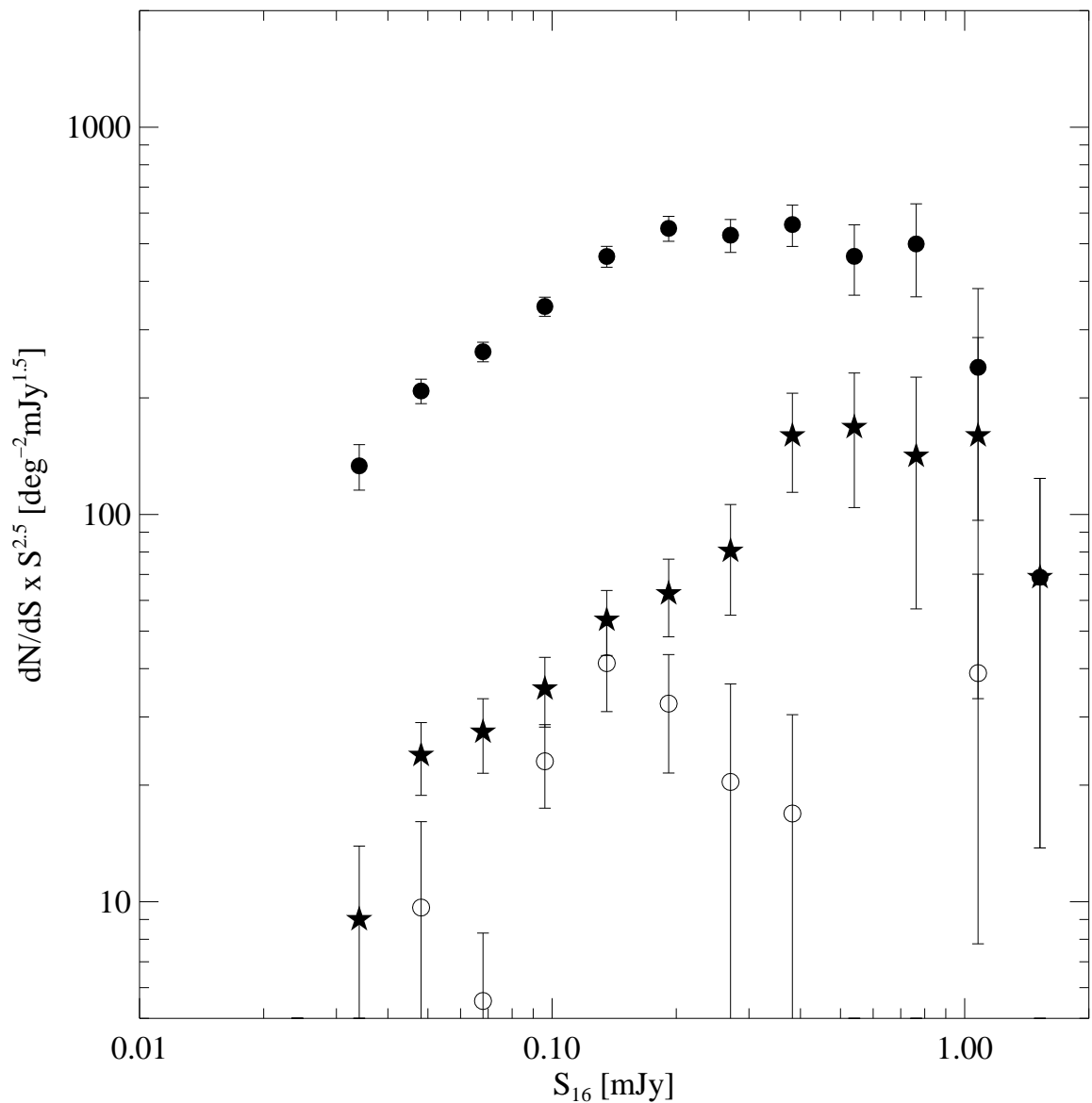} 
\caption{\label{fig: agnnc} Differential 16 \mic\ number counts for
 GOODS North and South combined ({\it filled circles}).  We show the
 contribution of sources identified as AGN ({\it star-shaped symbols}).  We
 also show the contribution of objects with 16 to 24 \mic\ flux
 ratios $>1.4$, and spectroscopic redshifts $1<z<2$, indicating possible silicate absorption ({\it open circles}). }
\end{figure}

In Figure \ref{fig: agnnc}, we show the contribution of AGN to the
number counts.  The result is largely similar to Figure \ref{fig:
 agnfrac}, with X-ray selected AGN dominating the AGN contribution at
bright fluxes, and the total AGN contribution decreasing at faint
fluxes.  We also show in the figure the contribution of 
silicate absorption candidate sources, which could be highly obscured
AGN.  These objects
occur mostly at faint flux levels, given their high redshift.

\subsection{Integrated Galaxy Light at 16 microns}
\label{sec: cirb}

Previous measurements of the monochromatic 15 \mic\ EBL have been
based on ISOCAM and AKARI data
\citep{2002A&A...384..848E,1999A&A...343L..65A}.
\cite{2003A&A...407..791M} inferred the contribution of ISO-detected
galaxies to the 15 \mic\ EBL to be $2.7\pm0.62$\ nW m$^{-2}$\
sr$^{-1}$\ at $S_{15}>30$\ \mujy\ by integrating the flux measured
from faint sources (the integrated galaxy light, IGL) including the
observations of lensing clusters. \cite{2010ApJ...716L..45H}\ improved
the measurement of the lensing cluster and determined an IGL value of
$1.9\pm0.5$\ nW m$^{-2}$\ sr$^{-1}$\ at $S_{15}>10$\ \mujy.

We can improve the measurement of the $\sim 15$\ \mic\ IGL using the
Spitzer sources. We have significantly more detections fainter
than 50 \mujy\ than any of the ISOCAM surveys, and we cover two
independent fields that were covered by the ultradeep AKARI measurement.  We first take the
average of the number counts for both GOODS fields, and combine them
with ISOCAM counts at fluxes $>1$\ mJy \citep{1999A&A...351L..37E,2002MNRAS.335..831G}.
Next, following \cite{2002A&A...384..848E}, we integrate $dIGL/dS$\
(defined by their Equation 6) over the range 30 \mujy\ to 1 mJy, after
fitting the counts with a $3^{rd}$\ degree polynomial (the dotted line in Figure \ref{fig: diffcounts}). 
A conservative
estimate of the uncertainty on $IGL_{15}$\ was obtained by fitting and integrating the
Poissonian upper and lower $1\sigma$\ uncertainty on the counts.

\begin{figure}[t!]
\plotone{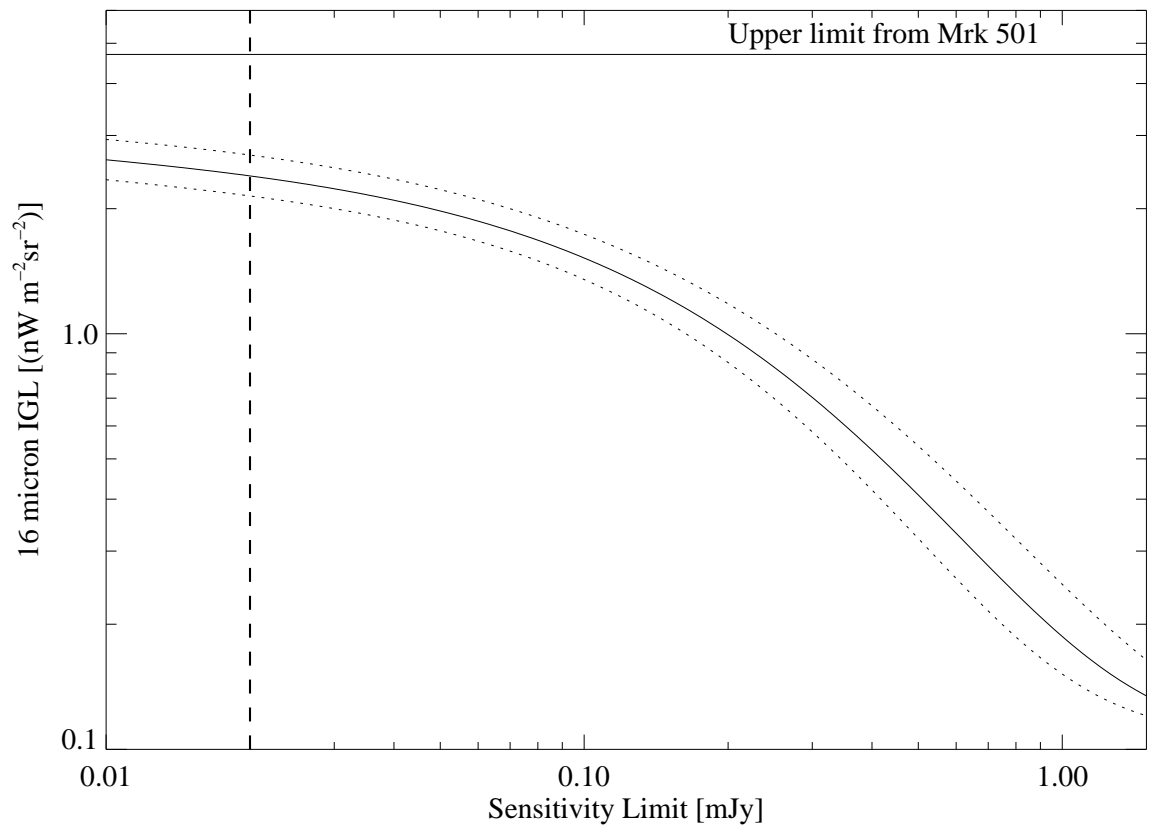}
\caption{\label{fig: igl} The integrated galaxy light (IGL) as a function of
sensitivity limit.  The IGL is extrapolated below the UDF sensitivity
({\it dotted vertical line}) assuming constant faint-end slope.  The
uncertainty ({\it dotted curve}) is estimated by
calculating the IGL for the
$1\sigma$\ upper and lower limits to the number counts.  The upper
limit to the IGL is shown \citep{2001A&A...371..771R}\ as the {\it solid horizontal line}.}
\end{figure}

We find $IGL_{16}~(S_{16} \ge 30$\mujy$) = 2.2 \pm 0.2$\ nW m$^{-2}$\
sr$^{-1}$. Note that if we color-correct the Spitzer sources to the
effective wavelength (14.3 \mic) of the ISOCAM LW3 filter as in Figure
\ref{fig: isocam}, we would obtain $IGL_{15}~(S_{15} \ge 25$\mujy$) =
1.8 \pm 0.2$\ nW m$^{-2}$\ sr$^{-1}$, just outside the $1\sigma$\
uncertainties from the \cite{2003A&A...407..791M}\ result.
Figure \ref{fig: igl}\ shows the IGL as a function of sensitivity limit.
The upper limit on the EBL, 4.7 nW m$^{-2}$\ sr$^{-1}$\, is taken from
the \cite{2001A&A...371..771R} high energy $\gamma$-ray measurement of
Mrk 501. Comparing the upper limit to the EBL, following Metcalfe et
al., we find the 16 \mic\ counts down to 30 \mujy\ appear to be
resolving at least $\sim 50$\% of the monochromatic 16 \mic\ EBL.
Most likely the resolved fraction of the 15\,$\mu$EBL is even larger
since the IGL at the faintest flux limits probed by the {\it Spitzer}
surveys (Figure 15) appears to be asymptoting towards a value of $\sim$3\ nW m$^{-2}$
if extrapolated to zero flux.

In principle, the 16 and 24 \mic\ IGL should be similar as they are
sampling mostly similar populations. Though, the 24 \mic\ band is more
sensitive to higher redshifts and, as we have shown, the 16 \mic\ band
picks up a slightly higher fraction of AGN. Our value is lower
than the quoted estimate of the 24 \mic\ IGL obtained by
\cite{2004ApJS..154...70P} of $2.7^{+1.1}_{-0.7}$\ nW m$^{-2}$\
sr$^{-1}$. However, that result was obtained by measuring the 24 \mic\
number counts down to 60 \mujy\ and extrapolating to fainter fluxes
using a fit to the faint-end slope. If we extrapolate the 16 \mic\ source
counts to fluxes 10 times fainter (down to 3 \mujy), assuming faint-end slope $\alpha=-1.9$, 
we obtain
$IGL_{16}~\sim 2.9$\ nW m$^{-2}$\ sr$^{-1}$, in good agreement with
the MIPS value.

\cite{2002A&A...383..838F}\ find that AGN contribute $\sim 17$\% of
the 15 \mic\ background \citep[see also][]{2007A&A...472..797L}.
Similarly, \cite{2006A&A...451..443M}\ place a lower limit of the AGN
contribution to the 15 \mic\ IGL of 4--10\%, using only
optically-selected AGN.  We measure the IGL for the AGN among the 16
\mic\ GOODS sources, and find a contribution to the 16 \mic\ IGL of $\sim15$\%
from X-ray and power-law selected AGN. 

Spitzer measurements at 8 and 24 \mic\ have typically determined a
marginally smaller contribution to the IGL from AGN.
\cite{2007ApJ...660..988B}\ and \cite{2006ApJ...642..126B}\ find
contribution of $\sim$10\% to the 24 \mic\ background based on X-ray
selection of AGN.  Most of these surveys also rely on X-ray selection of
AGN, and \cite{2006ApJ...642..126B}\ suggest that the AGN contribution
be corrected upwards by a factor of 1.5 to account for Compton-thick
sources. \cite{2006ApJ...644..143B}\ suggest that only 3--7\% of the
24 \mic\ 
background results from AGN, and \cite{2005AJ....129.2074F}\ place
the contribution at 10--15\%\ based on optical selection.
\cite{2004MNRAS.355..973S}\ note that the fractional AGN contribution
is, itself, an upper limit to the contribution to the IGL from AGN
radiation because significant IR light from these sources
arises from the host galaxy.

\section{Summary}

We have presented source catalogs for Spitzer 16 \mic\ observations of
the GOODS fields using the Peak-up Imaging capability.  We
surveyed the ACS area (150 sq. arcminutes) to depths of
40 and 65 \mujy\ (50\% completeness) in the Northern and Southern fields,
respectively.  In the $\sim$\ 10 sq. arcmin of the UDF, we reach 
30 \mujy.  We detect 840 (N) and 476 (S) objects.  These
observations are the widest contiguous-area PUI taken during the Spitzer mission,
and among the deepest.  The catalog will be available through the
NASA/IPAC Infrared Science Archive
(IRSA\footnote{http://irsa.ipac.caltech.edu}).

We validate the photometry by demonstrating that the Spitzer SEDs of
stars in the fields are consistent with standard templates from
Spitzer spectroscopy.  We compare PUI photometry with observations
within GOODS North by ISOCAM and find reasonable agreement after color
correction, though the uncertainties are large.  Comparison with AKARI
shows marginally significant disagreement ($\sim 30$\%), with Spitzer fluxes being
higher.  We match PUI sources within 1 arcsecond with Spitzer, Chandra
and HST detections.  We also report spectroscopic redshifts from the
literature where available.  The distribution of redshifts is similar
to that for 24 \mic\ sources despite the shallower MIR depth,
suggesting that redshifts so far available are mostly limited by
optical faintness.  We flag sources in the catalog that may have
quality issues: possible source confusion, bad concentration index,
lower than expected signal-to-noise ratio, low coverage.

Preliminary analysis of the survey find the following:

\begin{itemize}

\item Matching 16 \mic\ sources with Chandra hard-band detections
 finds that $\sim 11$\% of PUI sources have X-ray counterparts.  Combining
 these sources with power-law MIR selection for
 obscured AGN finds that about 15\% of objects in the catalog
 are potentially AGN. The fraction of AGN increases with increasing 16 \mic\
 flux density.  

\item The 16 to 24 \mic\ ratio shows significant variation with
 redshift.  Most sources lie within the locus expected for starbursts,
 IR luminous galaxies, and AGN.  A ratio $>1.4$\ appears likely to
 select predominantly sources at $1.1<z<1.6$, where the  the minimum
 between PAH emission peaks (as well as the 9.7 \mic\ silicate
 absorption) redshifts into the 24 \mic\ passband.  About 5\% of 16 \mic\ sources
 meet this color selection, though few are selected as AGN by X-ray
 emission or power-law SEDs and so any AGN contribution must be heavily
obscured.

\item Galaxy number counts show good agreement with previous
 surveys at similar wavelengths (from ISOCAM and AKARI).  The large
 number of sources and the two fields provide improvements in both
 the Poissonian errors and the effects of cosmic variance.  AGN make
 substantial contribution to the number counts at bright fluxes.

\item We measure the integrated galaxy light at 16 \mic, as a lower
 limit on the contribution to the monochromatic IR background, to be
 $2.2\pm0.2$ nW m$^{-2}$\ sr$^{-1}$.  Extrapolating to fluxes 10
 times fainter than the survey limit raises this value to 2.9 nW
 m$^{-2}$\ sr$^{-1}$, or 75\% of the extragalactic background light.
The contribution of sources which host AGN to the 16 \mic\ EBL is
 $\sim15$\%, similar to the fraction found in 24
 \mic\ surveys.

\end{itemize}

Future 15 \mic\ results are expected from AKARI.  Significantly more
area has been covered, which will allow improved study of rare sources.  The
Spitzer archive also contains many shallow (though small) fields which
were obtained in parallel to spectroscopic observations (Fajardo-Acosta
et al. 2010, in prep.).  Confusion-limited 16 \mic\ observations were
obtained by Spitzer in a small number of fields (e.g.\ Program ID=499;
PI=Colbert).   Finally, the recently launched WISE mission will include
12 \mic\ images over the whole sky

 \acknowledgments

This work is based in part on observations made with the {\it Spitzer
  Space Telescope}, which is operated by the Jet Propulsion
Laboratory, California Institute of Technology under NASA contract
1407. Support for this work was provided by NASA through an award
issued by JPL/Caltech. This research has made use of the NASA/ IPAC
Infrared Science Archive (IRSA), which is operated by the Jet Propulsion
Laboratory, California Institute of Technology, under contract with
the National Aeronautics and Space Administration.

{\rotate
\begin{center}
\begin{deluxetable}{lccccccccccccccc}
\tablecolumns{16} 
\tablewidth{0pc} 
\tablecaption{GOODS-N, Spitzer data} 
\tabletypesize{\scriptsize}
\tablehead{ 
\colhead{ID}    &  
\colhead{RA}   & 
\colhead{DEC}    &
\colhead{Cov\tablenotemark{a}}    &  
\colhead{$f_{ch1}$}    &  
\colhead{$\sigma_{ch1}$}    &  
\colhead{$f_{ch2}$}    &  
\colhead{$\sigma_{ch2}$}    &  
\colhead{$f_{ch3}$}    &  
\colhead{$\sigma_{ch3}$}    &  
\colhead{$f_{ch4}$}    &  
\colhead{$\sigma_{ch4}$}    &  
\colhead{$f_{16}$}    &  
\colhead{$\sigma_{16}$}    &  
\colhead{$f_{24}$}    &  
\colhead{$\sigma_{24}$}    \\
\colhead{}    &  
\colhead{(deg)}   & 
\colhead{(deg)}    &
\colhead{}    &  
\colhead{($\mu$Jy)}    &  
\colhead{($\mu$Jy)}    &  
\colhead{($\mu$Jy)}    &  
\colhead{($\mu$Jy)}    &  
\colhead{($\mu$Jy)}    &  
\colhead{($\mu$Jy)}    &  
\colhead{($\mu$Jy)}    &  
\colhead{($\mu$Jy)}    &  
\colhead{($\mu$Jy)}    &  
\colhead{($\mu$Jy)}    &  
\colhead{($\mu$Jy)}    &  
\colhead{($\mu$Jy)}    \\
\colhead{(1)}    &  
\colhead{(2)}   & 
\colhead{(3)}    &
\colhead{(4)}    &  
\colhead{(5)}   & 
\colhead{(6)}    &
\colhead{(7)}    &  
\colhead{(8)}    &  
\colhead{(9)}   & 
\colhead{(10)}    &
\colhead{(11)}    &  
\colhead{(12)}    &  
\colhead{(13)}   & 
\colhead{(14)}    &
\colhead{(15)}    &  
\colhead{(16)}   \\
}
\startdata
         0 &      189.319641 &       62.390766 &         11 &      84.80 &       4.24 &      60.40 &       3.02 &      55.20 &       2.94 &      40.20 &       2.16 &      145.1 &       11.3 &      225.0 &        5.5 \\
         1 &      189.310349 &       62.388977 &         12 &      11.50 &       0.58 &      14.50 &       0.73 &      19.00 &       1.50 &      12.90 &       0.99 &       55.3 &       10.4 &      200.0 &        5.7 \\
         2 &      189.326126 &       62.383057 &         20 &      38.30 &       1.92 &      49.10 &       2.46 &      55.30 &       2.83 &      39.30 &       2.06 &      107.2 &       16.5 &      483.0 &        5.7 \\
         3 &      189.294266 &       62.376274 &         20 &      40.00 &       2.00 &      47.50 &       2.38 &      39.60 &       2.11 &      44.20 &       2.30 &      240.4 &       11.5 &      386.0 &        4.9 \\
         4 &      189.330811 &       62.375191 &         24 &      10.10 &       0.51 &       6.98 &       0.36 &       7.48 &       0.63 &       6.27 &       0.63 &       53.9 &       11.5 &       26.6 &        5.6 \\
         5 &      189.284683 &       62.377537 &         15 &      33.50 &       1.68 &      30.20 &       1.51 &      21.00 &       1.40 &      20.50 &       1.24 &       96.0 &       18.3 &       89.1 &        6.2 \\
         6 &      189.301178 &       62.375061 &         21 &      10.90 &       0.55 &      10.50 &       0.53 &       8.67 &       0.75 &      11.60 &       0.82 &       75.5 &       11.7 &       60.5 &        3.6 \\
         7 &      189.317520 &       62.370590 &         27 &      28.40 &       1.42 &      24.10 &       1.21 &      20.70 &       1.15 &      34.00 &       1.78 &      129.3 &       14.6 &      133.0 &        4.9 \\
         8 &      189.314499 &       62.371925 &         26 &       1.16 &       0.08 &       1.13 &       0.10 &       0.47 &       0.52 &       3.13 &       0.56 &       56.7 &       11.3 &       50.7 &        4.0 \\
         9 &      189.306412 &       62.371181 &         22 &      50.70 &       2.54 &      37.50 &       1.88 &      32.20 &       1.70 &      30.10 &       1.60 &      339.4 &       13.7 &      436.0 &        5.8 \\
        10 &      189.266174 &       62.368641 &         19 &       7.67 &       0.39 &       5.57 &       0.30 &       4.81 &       0.84 &       4.93 &       0.72 &       62.8 &       15.0 &       18.3 &        4.6 \\
        11 &      189.292343 &       62.368526 &         18 &      32.10 &       1.61 &      29.90 &       1.50 &      21.80 &       1.23 &      23.50 &       1.30 &      130.8 &       12.8 &      125.0 &        4.0 \\
        12 &      189.351440 &       62.366024 &         21 &     242.00 &      12.10 &     199.00 &       9.95 &     131.00 &       6.57 &     214.00 &      10.71 &      365.8 &       17.6 &      395.0 &        7.7 \\
        13 &      189.351089 &       62.363888 &         21 &      18.80 &       0.94 &      14.00 &       0.70 &      12.40 &       0.77 &      14.00 &       0.87 &       62.8 &       11.8 &       60.5 &        6.2 \\
        14 &      189.281219 &       62.363293 &         15 &      47.40 &       2.37 &      51.00 &       2.55 &      52.10 &       2.66 &      59.60 &       3.03 &      129.2 &       16.5 &      240.0 &        3.4 \\
        15 &      189.276138 &       62.360157 &         22 &     101.00 &       5.05 &     103.00 &       5.15 &     117.00 &       5.88 &     123.00 &       6.17 &      219.1 &       14.1 &      303.0 &        5.2 
\enddata
\tablenotetext{(a)}{Coverage in number of exposures (30 seconds integration per exposure).}
\end{deluxetable}                       
\end{center}                        
}

{\rotate
\begin{center}
\begin{deluxetable}{cccccccccccccccccccc}
\tablecolumns{20} 
\tablewidth{0pc} 
\tablecaption{GOODS-N, ancillary data} 
\tabletypesize{\scriptsize}
\tablehead{ 
\colhead{$z_{\rm spec}$}   & 
\colhead{Ref.\tablenotemark{1}}    &
\colhead{X\tablenotemark{2}}    &  
\colhead{SB flux\tablenotemark{3}}    &  
\colhead{HB flux\tablenotemark{3}}    &  
\colhead{S\tablenotemark{4}}    &  
\colhead{Ratio\tablenotemark{5}}    &  
\colhead{N$_{cl}$\tablenotemark{6}}    &  
\colhead{N$_{cl16}$\tablenotemark{7}}    &  
\colhead{Q\tablenotemark{8}}    &  
\colhead{E\tablenotemark{9}}    &  
\colhead{$B$\tablenotemark{10}}    &  
\colhead{$V$\tablenotemark{10}}    &  
\colhead{$I$\tablenotemark{10}}    &  
\colhead{$z$\tablenotemark{10}}    &
\colhead{$\sigma_B$\tablenotemark{10}}    &  
\colhead{$\sigma_V$\tablenotemark{10}}    &  
\colhead{$\sigma_I$\tablenotemark{10}}    &  
\colhead{$\sigma_z$\tablenotemark{10}}    &  
\colhead{Icls\tablenotemark{11}}    \\
\colhead{}    &  
\colhead{}   & 
\colhead{}    &
\multicolumn{2}{c}{($10^{-15}$mW m$^{-2}$)}    &  
\colhead{}    &  
\colhead{}    &  
\colhead{}    &  
\colhead{}    &  
\colhead{}    &  
\colhead{}    &  
\multicolumn{8}{c}{(mag)}    &  
\colhead{}    \\
\colhead{(17)}    &  
\colhead{(18)}   & 
\colhead{(19)}    &
\colhead{(20)}    &  
\colhead{(21)}   & 
\colhead{(22)}    &
\colhead{(23)}    &  
\colhead{(24)}    &  
\colhead{(25)}   & 
\colhead{(26)}    &
\colhead{(27)}    &  
\colhead{(28)}    &  
\colhead{(29)}   & 
\colhead{(30)}    &
\colhead{(31)}    &  
\colhead{(32)}    &  
\colhead{(33)}    &  
\colhead{(34)}    &  
\colhead{(35)}    &  
\colhead{(36)}   \\
}
\startdata
    0.5820 &    72 &     0 &      0.000 &      0.000 &     0 &       2.32 &          1 &          1 &          1 &          0 &    -999.00 &    -999.00 &    -999.00 &    -999.00 &    -999.00 &    -999.00 &    -999.00 &    -999.00 &          0 \\
  -99.0000 &    -1 &     0 &      0.000 &      0.000 &     0 &       1.24 &          1 &          1 &          3 &          0 &      25.58 &      26.24 &      25.45 &      24.75 &       0.08 &       0.28 &       0.16 &       0.09 &          1 \\
  -99.0000 &    -1 &     0 &      0.000 &      0.000 &     0 &       3.09 &          1 &          1 &          1 &          0 &      26.85 &      26.14 &      25.80 &      25.01 &       1.45 &       0.19 &       0.17 &       0.09 &          1 \\
    1.5220 &    73 &     0 &      0.000 &      0.000 &     0 &       2.65 &          2 &          2 &          5 &          0 &      25.11 &      24.43 &      23.62 &      23.09 &       0.09 &       0.08 &       0.05 &       0.03 &          1 \\
    0.6760 &    72 &     0 &      0.000 &      0.000 &     0 &       1.76 &          1 &          1 &          9 &          0 &      24.71 &      23.68 &      22.82 &      22.59 &       0.04 &       0.03 &       0.02 &       0.01 &          1 \\
  -99.0000 &    -1 &     0 &      0.000 &      0.000 &     0 &       2.65 &          1 &          1 &          1 &          0 &      26.26 &      24.98 &      24.00 &      23.00 &       0.23 &       0.11 &       0.06 &       0.03 &          1 \\
  -99.0000 &    -1 &     0 &      0.000 &      0.000 &     0 &       2.04 &          1 &          1 &          1 &          0 &      27.90 &      27.02 &      25.50 &      24.44 &       0.50 &       0.34 &       0.11 &       0.05 &          1 \\
    0.5045 &    21 &     0 &      0.000 &      0.000 &     0 &       3.09 &          1 &          1 &          1 &          0 &      23.26 &      22.19 &      21.49 &      21.17 &       0.01 &       0.01 &       0.01 &       0.01 &          1 \\
  -99.0000 &    -1 &     0 &      0.000 &      0.000 &     0 &       3.72 &          1 &          1 &          1 &          0 &      25.82 &      25.51 &      25.04 &      24.54 &       0.08 &       0.06 &       0.05 &       0.04 &          1 \\
    0.9746 &    21 &     0 &      0.000 &      0.000 &     0 &       2.16 &          1 &          1 &          1 &          0 &      23.98 &      23.14 &      22.16 &      21.65 &       0.03 &       0.02 &       0.01 &       0.01 &          1 \\
    1.0084 &    21 &     0 &      0.000 &      0.000 &     0 &       3.61 &          1 &          1 &          9 &          0 &      24.38 &      23.96 &      23.27 &      22.90 &       0.02 &       0.02 &       0.01 &       0.01 &          1 \\
    1.3644 &    21 &     0 &      0.000 &      0.000 &     0 &       2.78 &          1 &          1 &          1 &          0 &      24.24 &      23.82 &      23.09 &      22.45 &       0.04 &       0.03 &       0.02 &       0.01 &          1 \\
    0.2810 &    21 &     0 &      0.000 &      0.000 &     0 &       3.27 &          1 &          2 &          5 &          0 &      20.70 &      19.13 &      18.39 &      18.07 &       0.01 &       0.01 &       0.01 &       0.01 &          1 \\
    0.8615 &    21 &     0 &      0.000 &      0.000 &     0 &       5.14 &          2 &          2 &          7 &          0 &    -999.00 &    -999.00 &    -999.00 &    -999.00 &    -999.00 &    -999.00 &    -999.00 &    -999.00 &          0 \\
    1.4498 &    21 &     1 &      8.270 &     14.120 &     0 &       2.14 &          1 &          1 &          1 &          0 &      22.97 &      22.77 &      22.37 &      21.98 &       0.01 &       0.01 &       0.01 &       0.01 &          1 \\
    0.9032 &    21 &     1 &      5.760 &      7.250 &     0 &       3.32 &          2 &          1 &          5 &          0 &      22.61 &      21.84 &      21.31 &      20.91 &       0.01 &       0.01 &       0.01 &       0.01 &          1 
\enddata
\tablenotetext{(1)}{$-1$\ no redshift;  
$10$\ -- \cite{2000ApJ...538...29C,2001AJ....121.2895C}; $12$\ -- \cite{2001AJ....122..598D};
$21$\ -- \cite{2004AJ....127.3121W}; $42$\ -- \cite{2005ApJ...622..772C},\cite{2004ApJ...617...64S}; 
$45$\ -- \cite{2008ApJ...675.1171P},\cite{2009ApJ...698.1380M},Spitzer IRS redshifts;
$51$\ -- \cite{2005ApJ...633..174T}; $61$\ -- \cite{2006ApJ...653.1004R};
$72$\ -- Stern et al.\ (in prep.); $73$\ -- \cite{2008ApJ...673L..21D};
$81$\ -- \cite{2008ApJ...689..687B}}
\tablenotetext{2}{Xray detection (1=detected, 0=not detected)}
\tablenotetext{3}{Alexander et al. (2003)}
\tablenotetext{4}{Star flag (1=star, 0=galaxy)}
\tablenotetext{5}{Concentration (ratio of apflux in 6:2 pixel radii)}
\tablenotetext{6}{Number of ch4 sources within 4 arcsec radius}
\tablenotetext{7}{Number of catalog sources within 5.4 arcsec radius}
\tablenotetext{8}{Quality Flag is assigned Bitwise:    Bit~0 $=$ included in catalog;      Bit~1 $=$ more than one 16 micron source within 5.4 arcsec; possible confusion;      Bit~2 $=$ source concentration differs from that expected for point source;      Bit~3 $=$ S/N ratio $< 5$, but coverage indicates it should be higher}
\tablenotetext{9}{Extend Flag (1=extended, 0=point source)}
\tablenotetext{10}{Giavalisco et al. (2004)}
\tablenotetext{11}{Number of i-band sources within 1 arcsec radius}
\end{deluxetable}                       
\end{center}                        
}

{\rotate
\begin{center}
\begin{deluxetable}{lccccccccccccccc}
\tablecolumns{16} 
\tablewidth{0pc} 
\tablecaption{GOODS-S, Spitzer data} 
\tabletypesize{\scriptsize}
\tablehead{ 
\colhead{ID}    &  
\colhead{RA}   & 
\colhead{DEC}    &
\colhead{Cov\tablenotemark{a}}    &  
\colhead{$f_{ch1}$}    &  
\colhead{$\sigma_{ch1}$}    &  
\colhead{$f_{ch2}$}    &  
\colhead{$\sigma_{ch2}$}    &  
\colhead{$f_{ch3}$}    &  
\colhead{$\sigma_{ch3}$}    &  
\colhead{$f_{ch4}$}    &  
\colhead{$\sigma_{ch4}$}    &  
\colhead{$f_{16}$}    &  
\colhead{$\sigma_{16}$}    &  
\colhead{$f_{24}$}    &  
\colhead{$\sigma_{24}$}    \\
\colhead{}    &  
\colhead{(deg)}   & 
\colhead{(deg)}    &
\colhead{}    &  
\colhead{($\mu$Jy)}    &  
\colhead{($\mu$Jy)}    &  
\colhead{($\mu$Jy)}    &  
\colhead{($\mu$Jy)}    &  
\colhead{($\mu$Jy)}    &  
\colhead{($\mu$Jy)}    &  
\colhead{($\mu$Jy)}    &  
\colhead{($\mu$Jy)}    &  
\colhead{($\mu$Jy)}    &  
\colhead{($\mu$Jy)}    &  
\colhead{($\mu$Jy)}    &  
\colhead{($\mu$Jy)}    \\
\colhead{(1)}    &  
\colhead{(2)}   & 
\colhead{(3)}    &
\colhead{(4)}    &  
\colhead{(5)}   & 
\colhead{(6)}    &
\colhead{(7)}    &  
\colhead{(8)}    &  
\colhead{(9)}   & 
\colhead{(10)}    &
\colhead{(11)}    &  
\colhead{(12)}    &  
\colhead{(13)}   & 
\colhead{(14)}    &
\colhead{(15)}    &  
\colhead{(16)}   \\
}
\startdata
         0 &       53.103943 &    -27.663679 &          2 &      29.60 &       1.48 &      20.80 &       1.05 &      23.70 &       1.37 &      17.00 &       1.10 &      120.5 &       23.7 &      120.0 &        5.7 \\
         1 &       53.142094 &    -27.664965 &          2 &      33.70 &       1.69 &      26.00 &       1.30 &      21.40 &       1.19 &      22.30 &       1.24 &      252.7 &       25.3 &      161.0 &        8.8 \\
         2 &       53.144711 &    -27.666069 &          3 &     171.00 &       8.55 &     149.00 &       7.45 &     112.00 &       5.62 &     576.00 &      28.81 &      830.6 &       39.8 &      938.0 &       11.5 \\
         3 &       53.110268 &    -27.667072 &          2 &      26.90 &       1.35 &      22.30 &       1.12 &      16.30 &       1.00 &      16.80 &       1.03 &      208.5 &       19.4 &      143.0 &        5.6 \\
         4 &       53.103607 &    -27.666225 &          4 &      21.10 &       1.06 &      19.70 &       0.99 &      13.80 &       0.92 &      13.00 &       0.90 &       81.8 &       14.6 &       44.2 &        4.9 \\
         5 &       53.102222 &    -27.669676 &          3 &      46.90 &       2.35 &      35.90 &       1.80 &      33.60 &       1.77 &      40.40 &       2.10 &      251.0 &       38.7 &      322.0 &        6.6 \\
         6 &       53.110622 &    -27.669182 &          3 &      31.10 &       1.56 &      26.90 &       1.35 &      20.50 &       1.17 &      17.30 &       1.04 &       99.2 &       17.9 &       74.0 &        5.4 \\
         7 &       53.123016 &    -27.669071 &          3 &     102.00 &       5.10 &      63.70 &       3.19 &      60.70 &       3.08 &      40.60 &       2.10 &      201.5 &       20.2 &      256.0 &        5.3 \\
         8 &       53.126549 &    -27.669138 &          3 &      25.90 &       1.30 &      26.20 &       1.31 &      19.70 &       1.11 &      19.90 &       1.14 &      174.7 &       15.9 &      118.0 &        5.1 \\
         9 &       53.129875 &    -27.671673 &          4 &      61.40 &       3.07 &      43.80 &       2.19 &      45.70 &       2.34 &      39.40 &       2.04 &      199.7 &       24.8 &      203.0 &        4.9 \\
        10 &       53.147671 &    -27.672302 &          4 &      44.70 &       2.24 &      39.10 &       1.96 &      27.80 &       1.47 &      25.60 &       1.38 &      158.2 &       24.8 &      108.0 &        4.9 \\
        11 &       53.102448 &    -27.672508 &          4 &      28.80 &       1.44 &      22.20 &       1.11 &      16.80 &       0.99 &      16.30 &       0.98 &      169.0 &       16.8 &      162.0 &        5.5 \\
        12 &       53.109512 &    -27.674276 &          4 &      28.50 &       1.43 &      22.10 &       1.11 &      19.20 &       1.08 &      22.10 &       1.23 &      180.3 &       17.4 &      175.0 &       11.9 \\
        13 &       53.114948 &    -27.674074 &          4 &      25.50 &       1.28 &      18.20 &       0.91 &      13.90 &       0.85 &      12.10 &       0.80 &      107.8 &       17.0 &       67.8 &        5.4 \\
        14 &       53.127071 &    -27.675257 &          4 &      33.80 &       1.69 &      32.10 &       1.61 &      25.40 &       1.35 &      19.10 &       1.09 &      116.0 &       16.5 &       57.1 &        3.9 \\
        15 &       53.090645 &    -27.675821 &          3 &      42.60 &       2.13 &      41.00 &       2.05 &      33.30 &       1.74 &      84.40 &       4.25 &      215.6 &       19.8 &      272.0 &        4.0 
\enddata
\tablenotetext{(a)}{Coverage in number of exposures (60 seconds per exposure).}
\end{deluxetable}                       
\end{center}                        
}

{\rotate
\begin{center}
\begin{deluxetable}{cccccccccccccccccccc}
\tablecolumns{20} 
\tablewidth{0pc} 
\tablecaption{GOODS-S, ancillary data} 
\tabletypesize{\scriptsize}
\tablehead{ 
\colhead{$z_{\rm spec}$}   & 
\colhead{Ref.\tablenotemark{1}}    &
\colhead{X\tablenotemark{2}}    &  
\colhead{SB flux\tablenotemark{3}}    &  
\colhead{HB flux\tablenotemark{3}}    &  
\colhead{S\tablenotemark{4}}    &  
\colhead{Ratio\tablenotemark{5}}    &  
\colhead{N$_{cl}$\tablenotemark{6}}    &  
\colhead{N$_{cl16}$\tablenotemark{7}}    &  
\colhead{Q\tablenotemark{8}}    &  
\colhead{E\tablenotemark{9}}    &  
\colhead{$B$\tablenotemark{10}}    &  
\colhead{$V$\tablenotemark{10}}    &  
\colhead{$I$\tablenotemark{10}}    &  
\colhead{$z$\tablenotemark{10}}    &
\colhead{$\sigma_B$\tablenotemark{10}}    &  
\colhead{$\sigma_V$\tablenotemark{10}}    &  
\colhead{$\sigma_I$\tablenotemark{10}}    &  
\colhead{$\sigma_z$\tablenotemark{10}}    &  
\colhead{Icls\tablenotemark{11}}    \\
\colhead{}    &  
\colhead{}   & 
\colhead{}    &
\multicolumn{2}{c}{($10^{-15}$mW m$^{-2}$)}    &  
\colhead{}    &  
\colhead{}    &  
\colhead{}    &  
\colhead{}    &  
\colhead{}    &  
\colhead{}    &  
\multicolumn{8}{c}{(mag)}    &  
\colhead{}    \\
\colhead{(17)}    &  
\colhead{(18)}   & 
\colhead{(19)}    &
\colhead{(20)}    &  
\colhead{(21)}   & 
\colhead{(22)}    &
\colhead{(23)}    &  
\colhead{(24)}    &  
\colhead{(25)}   & 
\colhead{(26)}    &
\colhead{(27)}    &  
\colhead{(28)}    &  
\colhead{(29)}   & 
\colhead{(30)}    &
\colhead{(31)}    &  
\colhead{(32)}    &  
\colhead{(33)}    &  
\colhead{(34)}    &  
\colhead{(35)}    &  
\colhead{(36)}   \\
}
\startdata
    0.6658 &    42 &     0 &      0.000 &      0.000 &     0 &       2.53 &          1 &          1 &         17 &          0 &      25.32 &      23.39 &      22.42 &      22.09 &       0.01 &       0.02 &       0.02 &       0.01 &          1 \\
    1.1120 &    30 &     0 &      0.000 &      0.000 &     0 &       3.31 &          2 &          1 &         27 &          0 &      23.07 &      22.54 &      21.96 &      21.52 &       0.01 &       0.01 &       0.01 &       0.01 &          1 \\
    0.2140 &    42 &     0 &      0.000 &      0.000 &     0 &       2.72 &          1 &          1 &          1 &          0 &      20.50 &      19.34 &      18.79 &      18.57 &       0.01 &       0.01 &       0.01 &       0.01 &          2 \\
    0.6243 &     1 &     0 &      0.000 &      0.000 &     0 &       2.49 &          1 &          1 &         17 &          0 &      26.43 &      24.76 &      23.77 &      23.06 &       0.01 &       0.05 &       0.04 &       0.02 &          1 \\
  -99.0000 &    -1 &     0 &      0.000 &      0.000 &     0 &       2.70 &          1 &          1 &          1 &          0 &      26.82 &      23.40 &      25.00 &      24.27 &       0.01 &       0.02 &       0.13 &       0.08 &          1 \\
  -99.0000 &    -1 &     1 &      0.000 &      1.300 &     0 &       2.58 &          2 &          1 &          3 &          0 &      24.53 &      23.66 &      22.55 &      21.91 &       0.04 &       0.02 &       0.01 &       0.01 &          1 \\
  -99.0000 &    -1 &     0 &      0.000 &      0.000 &     0 &       3.21 &          1 &          2 &          3 &          0 &    -999.00 &      27.25 &      27.28 &      25.01 &    -999.00 &       0.43 &       0.76 &       0.10 &          2 \\
    0.7343 &    42 &     0 &      0.000 &      0.000 &     0 &       2.30 &          1 &          1 &          1 &          0 &      23.46 &      22.25 &      21.08 &      20.64 &       0.03 &       0.01 &       0.01 &       0.01 &          1 \\
    1.3570 &    30 &     0 &      0.000 &      0.000 &     0 &       1.96 &          1 &          1 &          1 &          0 &      24.55 &      24.08 &      23.30 &      22.60 &       0.05 &       0.03 &       0.03 &       0.02 &          1 \\
    0.5659 &    42 &     0 &      0.000 &      0.000 &     0 &       2.53 &          1 &          1 &          1 &          0 &      22.93 &      21.69 &      20.86 &      20.52 &       0.02 &       0.01 &       0.01 &       0.01 &          1 \\
    1.1240 &    30 &     0 &      0.000 &      0.000 &     0 &       2.15 &          1 &          1 &          1 &          0 &      26.64 &      25.34 &      24.11 &      23.01 &       0.28 &       0.07 &       0.04 &       0.02 &          1 \\
    1.0368 &    42 &     0 &      0.000 &      0.000 &     0 &       2.88 &          1 &          1 &          1 &          0 &      25.21 &      24.09 &      23.14 &      22.48 &       0.09 &       0.02 &       0.02 &       0.01 &          1 \\
    1.0387 &    42 &     0 &      0.000 &      0.000 &     0 &       4.53 &          1 &          1 &          5 &          0 &      23.33 &      22.93 &      22.34 &      22.03 &       0.02 &       0.01 &       0.01 &       0.01 &          1 \\
  -99.0000 &    -1 &     0 &      0.000 &      0.000 &     0 &       2.15 &          1 &          1 &          1 &          0 &      23.68 &      23.23 &      22.46 &      21.92 &       0.04 &       0.02 &       0.02 &       0.01 &          1 \\
  -99.0000 &    -1 &     0 &      0.000 &      0.000 &     0 &       2.43 &          1 &          1 &          1 &          0 &      24.69 &      24.16 &      23.37 &      22.64 &       0.08 &       0.05 &       0.04 &       0.02 &          1 \\
    0.4260 &    72 &     0 &      0.000 &      0.000 &     0 &       2.85 &          1 &          1 &          1 &          0 &      22.06 &      20.98 &      20.47 &      20.20 &       0.01 &       0.01 &       0.01 &       0.01 &          1 
\enddata
\tablenotetext{1}{1 = LeFevre et al. (2004)
;      2 -- Szokoly et al. (2004)
;      3 -- Croom et al. (2001)
;      9 -- K20 Survey, Mignoli et al. (2005)
;     11 -- Strolger at al. (2004)
;     30 -- Vanzella et al. (2008)
;     41,42 -- \cite{2009A&A...494..443P},\cite{2010A&A...512A..12B}
;     62 -- Doherty et al. (2005)
;     70 -- Ravikumar et al. (2007)
;     72 -- Stern et al. (in prep.)
;     81 -- \cite{2008ApJ...677..219K}
}
\tablenotetext{2}{Xray detection (1=detected, 0=not detected)}
\tablenotetext{3}{     ~Luo et al. (2003)}
\tablenotetext{4}{Star flag (1=star, 0=galaxy)}
\tablenotetext{5}{Concentration (ratio of apflux in 6:2 pixel radii)}
\tablenotetext{6}{Number of ch4 sources within 4 arcsec radius}
\tablenotetext{7}{Number of catalog sources within 5.4 arcsec radius}
\tablenotetext{8}{Quality Flag is assigned Bitwise:    Bit~0 $=$
  included in catalog;      Bit~1 $=$ more than one 16 micron source
  within 5.4 arcsec; possible confusion;      Bit~2 $=$ source
  concentration differs from that expected for point source;
  Bit~3 $=$ S/N ratio $< 5$, but coverage indicates it should be
  higher;    Bit~4 $=$ source with coverage of two exposures instead of three }
\tablenotetext{9}{Extend Flag (1=extended, 0=point source)}
\tablenotetext{10}{Giavalisco et al. (2004)}
\tablenotetext{11}{Number of i-band sources within 1 arcsec radius}
\end{deluxetable}                       
\end{center}                        
}

\begin{deluxetable}{lrrrr}
\tablecaption{Numbers of Sources \label{tab: numbers}}
\tablehead{
\colhead{Field} &
\colhead{Sources} &
\colhead{Hard Xray} &
\colhead{other power-law AGN} &
\colhead{Blue ($f_{16}/f_{24}>1.4$)\tablenotemark{a}} 
}
\startdata
North & 840 & 96 & 9 & 55 (25)\\
South & 476 & 58 & 14 & 51 (34)
\enddata
\tablenotetext{a}{The number of sources with $f_{16}/f_{24}>1.4$\ and a $5\sigma$\ detection
at 24 \mic, with either
no spectroscopic redshift or a redshift $1.1<z<1.6$, indicating possible silicate
absorption.  The number in parenthesis indicates how many of these sources
have a quality flag value of 1.}

\end{deluxetable}

\clearpage

\begin{deluxetable}{lrrr}
\tablecaption{Sensitivities \label{tab: depth}}
\tablehead{
\colhead{Quantity} & 
\colhead{North} &
\colhead{South} &
\colhead{UDF} 
}
\startdata
Min Itime per pixel (s)\tablenotemark{a}      & 300 & 120 & 600 \\
Planned Itime per pixel (s)  & 600 & 120 & 960 \\
Mode Itime per pixel (s)     & 630 & 270 & 2100 \\
Min predicted depth\tablenotemark{b} (\mujy)  & 80  & 125 & 55 \\
Mode predicted depth (\mujy) & 48  & 80  & 30 \\
Mode measured Depth \tablenotemark{c} (\mujy)  & 45  & 80  & 30
\enddata
\tablenotetext{a}{Minimum exposure time per pixel for objects in the
 catalog}
\tablenotetext{b}{Minimum predicted $5\sigma$\ depth using the exposure time
calculator on the SSC website}
\tablenotetext{c}{Mode of measured $5\sigma$\ depth}

\end{deluxetable}

\clearpage

\begin{deluxetable}{lrrrrrr}
\tablecaption{GOODS-North Galaxy Number Counts \label{tab: nc north}}
\tablehead{
\colhead{$S_{low}$} & 
\colhead{$S_{high}$} & 
\colhead{$S_{avg}$} & 
\colhead{$N_{galaxies}$} &
\colhead{Observed $dN/dS$} &
\colhead{Corrected $dN/dS$} &
\colhead{Corrected $\delta~dN/dS$} \\
\colhead{(mJy)} &
\colhead{(mJy)} &
\colhead{(mJy)} &
\colhead{} &
\colhead{(deg$^{-2}$\ mJy$^{-1}$)} &
\colhead{(deg$^{-2}$\ mJy$^{-1}$)} &
\colhead{(deg$^{-2}$\ mJy$^{-1}$)} 
}
\startdata

    0.025  &     0.040   &    0.034   &       13  &   \nodata  &   \nodata  &   \nodata  \\
    0.040  &     0.056   &    0.048   &      106  &   3.0$\times 10^{5}$  &   4.5$\times 10^{5}$  &   1.6$\times 10^{4}$  \\
    0.080  &     0.112   &    0.096   &      151  &   1.1$\times 10^{5}$  &   1.3$\times 10^{5}$  &   9.6$\times 10^{3}$  \\
    0.112  &     0.159   &    0.136   &      128  &   6.6$\times 10^{4}$  &   7.1$\times 10^{4}$  &   6.0$\times 10^{3}$  \\
    0.159  &     0.224   &    0.192   &       93  &   3.4$\times 10^{4}$  &   3.4$\times 10^{4}$  &   3.5$\times 10^{3}$  \\
    0.224  &     0.317   &    0.271   &       61  &   1.6$\times 10^{4}$  &   1.6$\times 10^{4}$  &   2.0$\times 10^{3}$  \\
    0.317  &     0.448   &    0.382   &       45  &   8.2$\times 10^{3}$  &   8.2$\times 10^{3}$  &   1.4$\times 10^{3}$  \\
    0.448  &     0.632   &    0.540   &       22  &   2.8$\times 10^{3}$  &   2.8$\times 10^{3}$  &   7.4$\times 10^{2}$  \\
    0.632  &     0.893   &    0.763   &       11  &   1.0$\times 10^{3}$  &   1.0$\times 10^{3}$  &   4.0$\times 10^{2}$  \\
    0.893  &     1.262   &    1.078   &        3  &   1.9$\times 10^{2}$  &   1.9$\times 10^{2}$  &   1.6$\times 10^{2}$  \\
    1.262  &     1.783   &    1.522   &        2  &   9.1$\times 10^{1}$  &   9.1$\times 10^{1}$  &   7.3$\times 10^{1}$  \\
    1.783  &     2.518   &    2.150   &        0  &   \nodata  &   \nodata  &   \nodata  \\
    2.518  &     3.557   &    3.037   &        0  &   \nodata  &   \nodata  &   \nodata

\enddata
\end{deluxetable}

\clearpage

\begin{deluxetable}{lrrrrrr}
\tablecaption{GOODS-South Galaxy Number Counts \label{tab: nc south}}
\tablehead{
\colhead{$S_{low}$} & 
\colhead{$S_{high}$} & 
\colhead{$S_{avg}$} & 
\colhead{$N_{galaxies}$} &
\colhead{Observed $dN/dS$} &
\colhead{Corrected $dN/dS$} &
\colhead{Corrected$\delta~dN/dS$} \\
\colhead{(mJy)} &
\colhead{(mJy)} &
\colhead{(mJy)} &
\colhead{} &
\colhead{(deg$^{-2}$\ mJy$^{-1}$)} &
\colhead{(deg$^{-2}$\ mJy$^{-1}$)} &
\colhead{(deg$^{-2}$\ mJy$^{-1}$)} 
}
\startdata

    0.025   &    0.040  &     0.034   &       14  &   3.1$\times 10^{5}$  &   6.2$\times 10^{5}$  &   1.4$\times 10^{5}$  \\
    0.040   &    0.056  &     0.048   &       20  &   2.6$\times 10^{5}$  &   3.7$\times 10^{5}$  &   8.4$\times 10^{4}$  \\
    0.056   &    0.080  &     0.068   &       38  &   1.5$\times 10^{5}$  &   1.8$\times 10^{5}$  &   2.8$\times 10^{4}$  \\  
    0.080   &    0.112  &     0.096   &      107  &   9.2$\times 10^{4}$  &   1.2$\times 10^{5}$  &   1.0$\times 10^{4}$  \\
    0.112   &    0.159  &     0.136   &      109  &   5.9$\times 10^{4}$  &   6.6$\times 10^{4}$  &   6.0$\times 10^{3}$  \\
    0.159   &    0.224  &     0.192   &       83  &   3.2$\times 10^{4}$  &   3.4$\times 10^{4}$  &   3.6$\times 10^{3}$  \\
    0.224   &    0.317  &     0.271   &       45  &   1.2$\times 10^{4}$  &   1.2$\times 10^{4}$  &   2.1$\times 10^{3}$  \\
    0.317   &    0.448  &     0.382   &       21  &   4.0$\times 10^{3}$  &   4.0$\times 10^{3}$  &   1.1$\times 10^{3}$  \\
    0.448   &    0.632  &     0.540   &       11  &   1.5$\times 10^{3}$  &   1.5$\times 10^{3}$  &   6.0$\times 10^{2}$  \\  
    0.632   &    0.893  &     0.763   &       10  &   9.6$\times 10^{2}$  &   9.6$\times 10^{2}$  &   4.1$\times 10^{2}$  \\
    0.893   &    1.262  &     1.078   &        3  &   2.0$\times 10^{2}$  &   2.0$\times 10^{2}$  &   1.6$\times 10^{2}$  \\
    1.262   &    1.783  &     1.522   &        1  &   4.8$\times 10^{1}$  &   4.8$\times 10^{1}$  &   3.9$\times 10^{1}$  \\
    1.783   &    2.518  &     2.150   &        2  &   6.8$\times 10^{1}$  &   6.8$\times 10^{1}$  &   5.5$\times 10^{1}$  \\
    2.518   &    3.557  &     3.037   &        1  &   2.4$\times 10^{1}$  &   2.4$\times 10^{1}$  &   1.9$\times 10^{1}$

\enddata
\end{deluxetable}

\clearpage

\clearpage

\clearpage

\clearpage

\clearpage

\clearpage

\clearpage

\clearpage

\clearpage

\clearpage

\clearpage

\clearpage

\clearpage

\clearpage

\clearpage

\end{document}